\documentclass{emulateapj}

\usepackage{apjfonts}
\usepackage{amsmath}
\usepackage{amssymb}
\usepackage{natbib}
\usepackage{xspace}
\usepackage{graphicx}
\usepackage{rotating}

\newcommand{\Rsun}{\ensuremath{\mbox{R}_\odot}\xspace}
\newcommand{\Rstar}{\ensuremath{\mbox{R}_*}\xspace}

\newcommand{\Msun}{\ensuremath{M_\odot}\xspace}

\def\gx{\mbox{GX\,301$-$2}\xspace}

\def\kev{ke\kern -0.05em V\xspace}

\slugcomment{accepted ApJ}

\shorttitle{GX 301-2 with Suzaku}
\shortauthors{Suchy et al.}
 
\begin{document}
\title{Broadband spectroscopy using two \textsl{Suzaku} observations of the HMXB GX\,301$-$2 } 

\author{Slawomir~Suchy\altaffilmark{1},
Felix~F\"urst\altaffilmark{2},
Katja~Pottschmidt\altaffilmark{3,4},
Isabel~Caballero\altaffilmark{5},
Ingo~Kreykenbohm\altaffilmark{2},
J\"orn~Wilms\altaffilmark{2},
Alex~Markowitz\altaffilmark{1},
Richard~E.~Rothschild\altaffilmark{1}
}

\altaffiltext{1}{University of California, San Diego, Center for
  Astrophysics and Space Sciences, 9500 Gilman Dr., La Jolla, CA
  92093-0424, USA}
\altaffiltext{2}{Dr. Karl Remeis Sternwarte \& Erlangen Center for Astroparticle Physics, Sternwartstr. 7, 96049 Bamberg, Germany}
\altaffiltext{3}{Center for Space Science and Technology, University of Maryland Baltimore County, 1000 Hilltop Circle, Baltimore, MD 21250, USA}
\altaffiltext{4}{CRESST and NASA Goddard Space Flight Center, Astrophysics Science Division, Code 661, Greenbelt, MD 20771, USA}
\altaffiltext{5}{CEA Saclay, DSM/IRFU/SAp- UMR AIM (7158), CNRS/CEA, University Paris 7, Diderot, Gif sur Yvette, France}

\email{ssuchy@ucsd.edu}

\begin{abstract} 
We present the analysis of two \textsl{Suzaku} observations of \gx at two orbital phases after the periastron passage. 
Variations in the column density of the line-of-sight absorber are observed, consistent with accretion from a clumpy wind.
In addition to a CRSF, multiple  fluorescence emission lines were detected in both observations. 
The variations in the pulse profiles and the CRSF throughout the pulse phase have a signature of a magnetic dipole field. 
Using a simple dipole model we calculated the expected magnetic field values for different pulse phases and were able to 
extract a set of geometrical angles, loosely constraining the dipole geometry in the neutron star. From the variation of the CRSF width and energy, we found a geometrical solution for the dipole, making the inclination consistent with 
previously published values. 

\end{abstract}

\keywords{X-rays: stars --- X-rays: binaries --- stars: pulsars:
  individual (GX 301$-$2) --- stars: magnetic fields}

\section{Introduction}\label{sec:introduction}
The High Mass X-ray Binary (HMXB) system \gx was  discovered in 1969 April during a 
balloon experiment \citep{Lewin:1971, McClintock:1971}. 
The system consists of an accreting neutron star (NS) fed by 
the surrounding stellar wind of the B type emission line 
companion Wray 977 \citep{Jones:1974}. 
A recent luminosity estimate derived from atmospheric 
models puts its distance at $\sim 3$\,kpc \citep{Kaper:2006}, 
the value utilized in this paper. 
The orbital period was established to be $\sim 41$\,days \citep{White:1978} 
using \textsl{Ariel}~5 observations and was refined with the 
\textsl{Burst And Transient Source Experiment (BATSE)} to 
$\sim 41.5$\,days with an eccentricity of $\sim 0.46$ \citep{Koh:1997}. 
\citet{Doroshenko:2010a} discussed a possible orbital evolution and determined 
an orbital period of $41.482\pm0.001$\,d, assuming no change in orbital period. 
\citet{Kaper:2006} determined that the mass of the companion was 
in the range  $39\,\Msun < M < 53\,\Msun$ and the radius of Wray\,977 was  
$\Rstar \sim$62\,\Rsun, obtained by fitting atmosphere models.
 
\begin{figure}
 \plotone{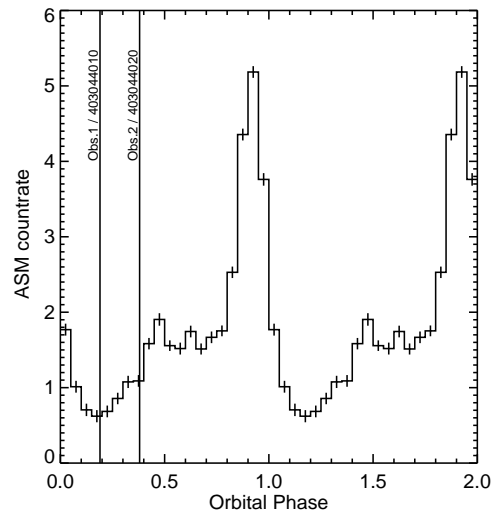}
\caption{\textsl{RXTE}/ASM 1.5$-$12\,keV light curve folded with the 
orbital period of 41.48\,days. The two lines indicate the times of the orbital phase when 
the \textsl{Suzaku} observations were performed. The profile shows the strong flux increase during the pre-periastrion 
flare at orbital phase $\sim 0.9$. For clarity, two binary orbits are shown.}
\label{fig:orbitlc}
\end{figure}

The X-ray flux is highly variable throughout an individual binary orbit 
but follows a distinct pattern when averaged over multiple orbits (see Figure~\ref{fig:orbitlc}). 
Shortly before the periastron passage, 
the X-ray luminosity increases drastically in the energy band above $\sim 5$\,keV, as 
seen in \textsl{Rossi X-ray Timing Explorer (RXTE)}/ All Sky Monitor (ASM) data \citep{Leahy:2002}.
The NS passes closest to the companion at a distance of $\sim 0.1\Rstar$ \citep{Pravdo:1995}.
Shortly after the periastron passage, $\phi_\text{orb} \sim 0.2$, the X-ray luminosity 
dips for a short period of time. \citet{Leahy:2002} demonstrated that neither a simple 
spherical wind model nor a circumstellar disk model around Wray 977
are sufficient to describe the observed variations in the folded 
\textsl{RXTE}/ASM data.
An additional stream component is able to account for the sudden increase in X-ray 
luminosity, as the NS passes trough the stream shortly before periastron and accretes more material. 
This model also explains a slightly higher X-ray luminosity around $\phi_\text{orb} \sim 0.5$, when the 
NS passes through the accretion stream a second time.

Pulsations with a period of $\sim 700$\,s were discovered in the 
\textsl{Ariel-5} observations \citep{White:1976}, making \gx
one of the slowest known pulsars. The pulse period has varied drastically 
throughout the last $\sim 20$ years \citep{Pravdo:2001, Evangelista:2010}.
Prior to 1984, the pulse period stayed relatively constant at 695\,s$-$700\,s 
and then spun up between 1985 and 1990 to $\sim 675$\,s. 
From 1993 until the beginning of 2008, the change in the spin reversed again, showing a decline. 
\textsl{Fermi}/Galactic Burst Monitor (GBM) data\footnote{http://www.batse.msfc.nasa.gov/gbm/science/} 
have revealed that \gx experienced  
another spin reversal and briefly spun-up with the pulse 
period decreasing from $\sim687$\,s to $\sim 681$\,s between 
May 2008 and October 2010. Since October 2010, the pulse period has shown only small variations around $\sim 681$\,s. 

Most recently, \citet{Gogus:2011} discovered a peculiar 1\,ks dip in the luminosity of \gx, where 
the pulsations disappeared for one spin cycle during the dip. Several such dips 
have been previously observed in Vela~X-1 \citep{Kreykenbohm:1999, Kreykenbohm:2008}, where it is assumed 
that the accretion on the NS was interrupted for a short period of time. 

The pulse phase average spectrum of \gx is described using a power law with a high energy cutoff. 
The continuum does not show a strong variation in the intrinsic parameters 
($\Gamma, E_\text{cut}$, and $E_\text{fold}$) throughout the orbit  
\citep{Mukherjee:2004}, as seen in two data sets from 
\textsl{RXTE},  taken in 1996 and 2000, 
sampling most phases of the binary orbit. 
One of the major characteristics of the 
X-ray spectrum of \gx is the high and strongly variable
column density of its line-of-sight absorber throughout the orbit, 
indicative of a clumpy stellar wind  ($N_\text{H} = 10^{22}-10^{24}$ cm$^{-2}$). 
In addition to the high column density, a very bright Fe K$\alpha$ emission line can be observed.
This line has shown a strong
correlation with the observed luminosity, indicating that the line is 
produced by local clumpy matter surrounding the neutron star \citep{Mukherjee:2004}. 
\citet{Kreykenbohm:2004} used the \textsl{RXTE} data set from 2000 to 
perform phase resolved spectroscopy and showed that an absorbed and partially 
covered pulsar continuum (power law with Fermi-Dirac cutoff) as well as a 
reflected and absorbed pulsar continuum were consistent with the data. 

A cyclotron resonance scattering feature (CRSF) at $\sim35$\,keV 
was first discovered with \textsl{Ginga} \citep{Mihara:1995}. 
\citet{Orlandini:2000} found systematic deviations from a power law continuum at $\sim 20$ and 
$\sim 40$\,keV in \textsl{BeppoSAX}, where the former could not be
confirmed as a CRSF due to the proximity of the continuum cutoff. 
\citet{Kreykenbohm:2004} excluded the existence of a CRSF 
at $\sim 20$\,keV and showed that the CRSF centroid 
energy varies between 30--38\,keV over the pulse rotation of the NS. 
Furthermore, they showed that the CRSF centroid energy and width 
are correlated. 

We report on two observations of \gx performed with the \textsl{Suzaku} 
satellite in mid 2008 and early 2009. This paper is structured as follows: Section 2 discusses 
the observations and data reduction. Section 3 shows the phase 
averaged results. Section 4 discusses the pulse profiles and 
phase resolved spectra. Sections 5 and 6 discuss the results 
and conclusions, respectively.

\section{Observation and Data Reduction}\label{data}
\textsl{Suzaku} observed \gx on 2008 August 25 with an exposure time of 
$\sim 10$\,ks (ObsID 403044010; hereafter Obs.~1). The observation was
cut short by a set of target of opportunity observations and was continued on 
2009 January 5, acquiring  an additional $\sim 60$\,ks exposure time 
(ObsID 403044020; Obs.~2). 
Both main instruments, the X-ray Imaging Spectrometer \citep[XIS;][]{Mitsuda:2007} 
and the Hard X-ray Detector \citep[HXD;][]{Takahashi:2007}
were used in these observations.  The two observations correspond to orbital 
phases of 0.19 (Obs.~1) and 0.38 (Obs.~2), where Obs.~1 falls into the 
lowest flux part of the binary orbit (see Figure~\ref{fig:orbitlc}). 
The $2-10$\,keV absorbed flux was $1.6\times10^{-10}$\,erg\,cm$^{-2}$\,s$^{-1}$ for Obs.~1 and 
$\sim 8 \times10^{-10}$\,erg\,cm$^{-2}$\,s$^{-1}$ for Obs.~2 (see Table 1 for details). 
Both observations were performed using the HXD nominal pointing to 
enhance the sensitivity of the HXD detectors.

The XIS detectors consist of two front illuminated (FI) CCDs XIS 0 and 3, and one back illuminated (BI) CCD, XIS 1. 
All three instruments are sensitive between $\sim 0.5$\,keV and $\sim 11$\,keV, 
although the two FI cameras have a higher sensitivity above $\sim 2$\,keV, and the 
BI chip is more sensitive below $\sim 2$\,keV. 
To minimize possible pile-up, the XIS instruments were operated with the 1/4 window option  with 
a readout time of 2\,s. Data were taken in both $3\times3$ and $5\times5$ editing modes, 
which were extracted individually  with the \textsl{Suzaku} FTOOLS version 16 as part of HEASOFT 6.9. 
The unfiltered XIS data were reprocessed with the most recent calibration files available and then screened with the 
standard selection criteria as described by the \textsl{Suzaku} ABC 
guide\footnote{http://heasarc.gsfc.nasa.gov/docs/suzaku/analysis/abc/}. The response matrices 
(RMFs) and effective areas (ARFs) were weighted according to the exposure times of the 
different editing modes. The XIS data were then grouped with the number of 
channels per energy bin corresponding to the half width half maximum of the spectral 
resolution, i.e. grouped by 8, 12, 14, 16, 18, 20, and 22 channels starting at 0.5, 1, 2, 3, 
4, 5, 6, and 7\,keV, respectively (M.~A.~Nowak, 2010, private communication). 
The XIS spectral data were used in the energy range of 
$2-10$\,keV for reasons described below. 

The HXD consists of two non-imaging instruments: 
the PIN silicon diodes (sensitive between 12--60\,keV) and the GSO/BGO phoswich 
counters (sensitive above $\sim 30$\,keV). To determine the PIN background, the  
\textsl{Suzaku} HXD team provides 
the tuned PIN Non X-ray Background (NXB) for each individual observation. 
In addition, the cosmic X-ray Background (CXB) was simulated following 
the example in the ABC guide and both backgrounds were added together. 
The PIN data were grouped by a factor of 5 below and by a factor of 10 above 50\,keV for Obs.~1. 
In Obs.~2 the channels were grouped by 3 throughout the whole energy range.
GSO data were extracted and binned following the \textsl{Suzaku} ABC guide.
The PIN data energy range of 15--60\,keV and the GSO data energy range of $50-90$\,keV were used 
for the phase averaged analysis. Due to the short exposure times, 
no GSO data were extracted in Obs.~1 and in the phase resolved analysis of Obs.~2.

\subsection{XIS Responses} 
\begin{figure}
 \plotone{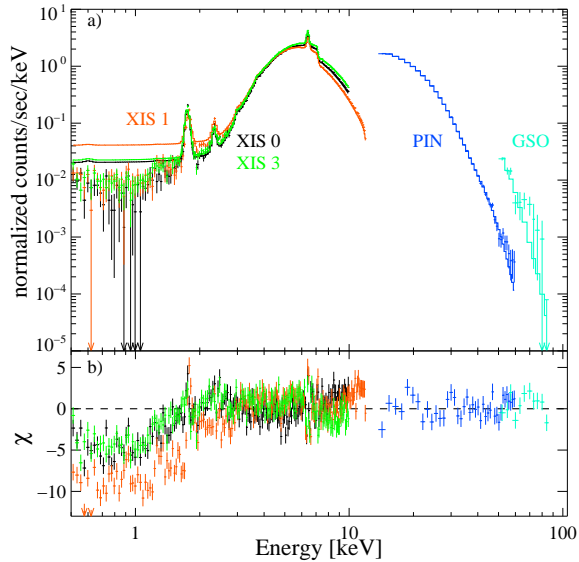}
\caption{(a) Best fit models of the \textsl{Suzaku} \gx second observation for the 3 individual XIS instruments, and the two HXD instruments. At lower energies, one clearly sees the constant level in the modeled flux discussed in the text. (b) shows that the residuals below $\sim 2$\,keV are very pronounced. }
\label{fig:badspec}
\end{figure}

\gx is a source with very large and highly variable photoelectric absorption in the line of sight as well as a very strong 
Fe K$\alpha$ emission line. For spectral modeling, the calculated responses
of the XIS detectors showed a `leakage' emission level \citep{Matsumoto:2006},
which stemmed from the instrument characteristics. 
That is, a small fraction of an event charge cloud is registered at energies below the peak energy, 
creating a 'low energy tail' in the spectrum. This `tail' is about three orders 
of magnitude weaker than the intensity of the peak count rate of the measured photon. 
The response to the Fe line is therefore described as a Gaussian plus a constant at 
lower energies, approximately three orders of magnitude below the 
peak value and extending to below 1\,keV. 
Due to the strong Fe~K$\alpha$ 
line and the very strong absorption in this source, the constant level response of the Fe line 
extends above the measured flux below 2\,keV. This results in significant residuals below $\sim2$ keV, which
seem to be stronger for the BI spectrum then for that of the FI CCD. 
Figure~\ref{fig:badspec} shows the phase averaged spectrum from the second observation of \gx and 
the best fit models using the HEASOFT generated XIS responses as detailed in section 3.

Together with the \textsl{Suzaku} Guest Observer Facility (GOF), which provided for us an experimental response 
parameter file where the `low energy tail' is ignored, we studied this behavior in more detail. 
This new response for all three instruments was applied to the individual XIS spectra. 
Due to the fact that the new response completely excluded 
the missing `tail', we now observed an excess of the data with respect to the model below $\sim 2\,$keV. The 
introduction of additional power law and/or black body components could improve the 
fits at lower energies. Due to the fact that the true level of the tail is not known, these additional components cannot be interpreted 
physically, and we decided to use the original response matrix and confine the selected 
XIS energy range to above 2\,keV.

\section{Phase averaged spectrum} 
\subsection{Spectral modeling}

Phase averaged broad band spectra in the 2--60\,keV (Obs.~1) 
and 2--90\,keV (Obs.~2) energy ranges were obtained.
From the technical description of the XIS 
instruments\footnote{http://heasarc.nasa.gov/docs/suzaku/prop\_tools/suzaku\_td/}, 
it is known that small discrepancies, 
e.g., in fitted power law slope, have been observed between FI and BI XIS instruments. 
A difference of $\sim0.05$ in the power law index has been observed in calibration data and 
was also observed here. When modeling the FI and BI instruments 
with a common power law index, the residuals of the BI instrument show a systematic deviation. 
However, only the FI or BI instruments could be modeled together with the HXD instruments. 
Due to this fact and the higher sensitivity of the FI XIS instruments above 2\,keV, 
we concentrated our discussion on the results obtained with the two FI instruments. 
Each data set, FI and BI,  was modeled individually with the HXD data to compare the differences in the 
continuum. We found that when using the FI and BI instruments individually with the HXD instruments, 
in both observations the best fit spectral values are consistent with each other within error bars, indicating that the 
usage of the HXD reduced the observed discrepancies in the power law parameters. 

\begin{figure}
 \plotone{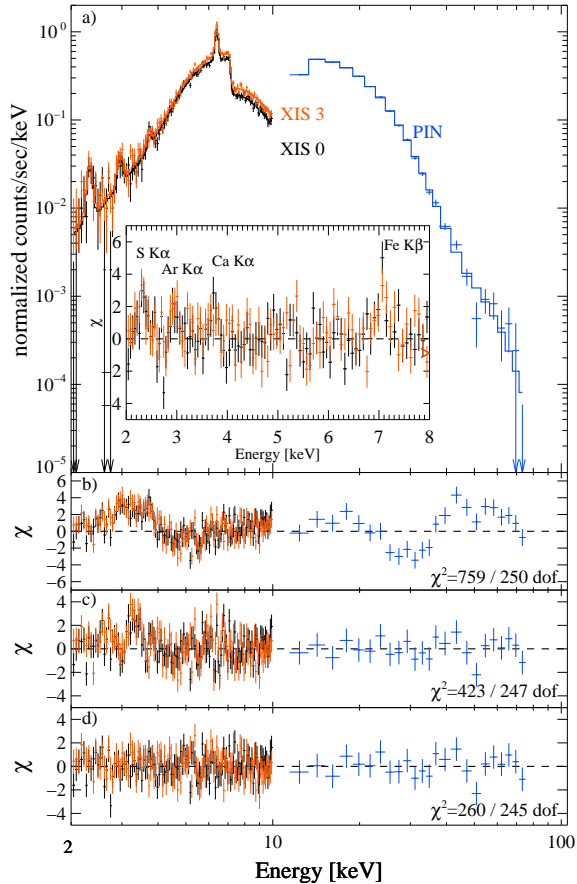}
\caption{Data and best fit model for Obs.~1. a) shows the data and the best fit model. The inset shows 
the residual for the different emission lines observed in XIS data, where the Fe K$\alpha$ line is already included in the model. b) shows the best fit residuals without partial covering and the CRSF. c) has only the CRSF added, and d) has both componenst included.}
\label{fig:spec010}
\end{figure}

\begin{figure}
 \plotone{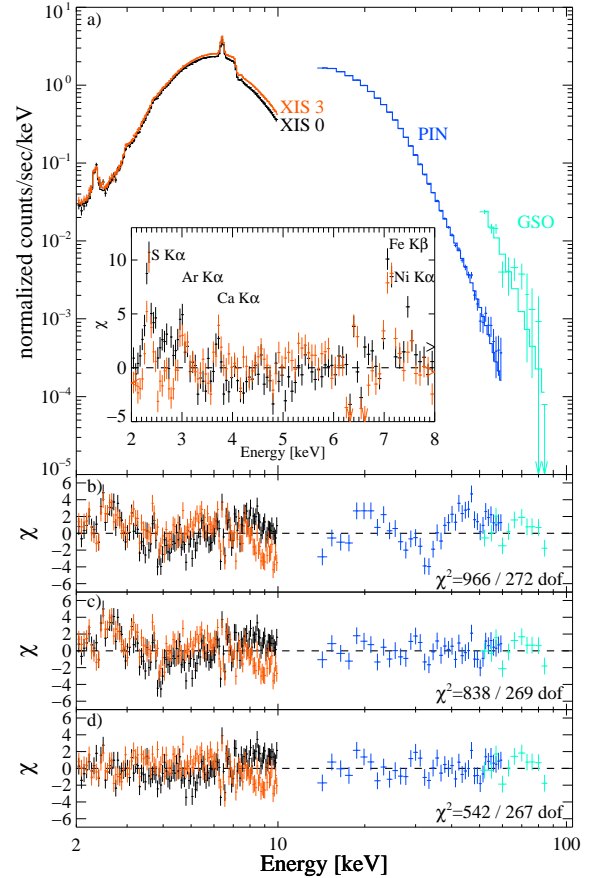}
\caption{Same as Figure~\ref{fig:spec010} for Obs.~2. In this case b) shows the residuals without partial covering and the CRSF, c) has only the CRSF added and 
d) has both components included.  }
\label{fig:spec020}
\end{figure}

The continuum model for pulsars can so far only be modeled with empirical models, consisting of a \texttt{powerlaw} 
with a cutoff at higher energies, which is typical for this kind of source \citep{Coburn:2001}. 
Three empirical models are widely used: the simple high energy cutoff (\texttt{highecut}) and 
the Fermi-Dirac cutoff \citep[\texttt{fdcut}, ][]{Tanaka:1986} both are used in combination with a simple power law component.
In addition, the negative-positive exponential powerlaw model \texttt{NPEX} \citep{Mihara:1995} 
is a slightly more complicated model including the power law component. 
For the data analyzed in this work, the best results have been obtained with the \texttt{fdcut} model: 
\begin{equation}
I_\text{cont} = A_\text{PL} \frac{ E^{-\Gamma}}{\text{exp}(E-E_\text{cut}/E_\text{fold})+1}, 
\end{equation}
where $A_\text{PL}$ is the normalization at 1\,keV, $E_\text{cut}$ is the 
cutoff energy, and $E_\text{fold}$ is the folding energy of the Fermi-Dirac cutoff. 
For the Obs.~1 and Obs.~2 FI data sets the best fit $\chi^2$'s were 260 and 542 with 
245 and 267 degrees of freedom (dof), respectively (see Figure~\ref{fig:spec010} 
and \ref{fig:spec020}). Replacing the smoother Fermi-Dirac cutoff with the \texttt{highecut} 
model in Obs.~2,resulted in slightly worse residuals ($\Delta \chi^2 \approx 15$) compared to the best fit results of the FD-cutoff model (Figure~\ref{fig:spec010} d) and an emission line-like residual at $\sim 35$\,keV, which is most likely due to the sharp break of the power law at the cutoff energy in this model.
In comparison, the NPEX model shows even worse residuals ($\Delta \chi^2 \approx 200$) and only results in reasonable fits
when the exponential curvature at higher energies is independent in the partially and fully covered component. 

\begin{figure}
   \plotone{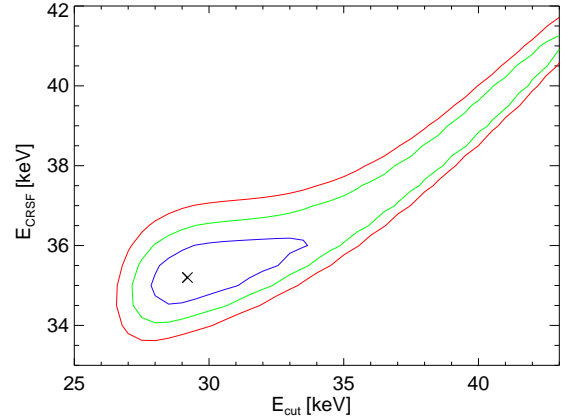}
\caption{Contour plots of $E_\text{CRSF}$ vs $E_\text{cut}$. The different contours indicate the 1, 2, and 3 sigma 
contours for the FI spectrum of Obs.~2. The X indicates the best fit values for both parameters: $E_\text{cut} = 29.2$\,keV and $E_\text{CRSF} = 35.2$\,keV.}
\label{fig:cont}
\end{figure}

The best fit  cutoff energy is very close to that of the observed CRSF feature (see below) and
these two parameters are rather strongly correlated (see Figure~\ref{fig:cont}). To avoid a degeneracy of the 
CRSF values due to a changing cutoff energy, $E_\text{cut}$ was frozen in all the fits to the best fit value from the Obs.~2 FI spectrum (29.2\,keV).

Previous observations of \gx indicated the existence of clumps in the stellar wind 
\citep{Kreykenbohm:2004, Mukherjee:2004}, which were modeled using partial 
covering absorption in addition to fully covered photoelectric absorption of the smooth stellar wind. 
In the present analysis, the low energy portion of the spectrum also required a partial covering component 
(Figure~\ref{fig:spec010}b and \ref{fig:spec020}b), which was modeled 
using the \texttt{TBnew} model (Wilms  et al., 2011, in prep.)\footnote{http://pulsar.sternwarte.uni-erlangen.de/wilms/research/tbabs/},
an updated version of the existing \texttt{TBabs} model \citep{wilms:2000}. 
In addition, a non-relativistic, optically-thin Compton scattering 
component \texttt{cabs} was included, as is necessary for column densities $N_\text{H} > 5 \times 10^{22}$ cm$^{-2}$,
where the plasma becomes Compton thick and part of the emission is scattered
out of the line of sight.
The $N_\text{H}$ values from \texttt{TBabs} and from the \texttt{cabs} component 
were set equal and treated as one model component for the smooth stellar wind 
($N_\text{H,1}$ = \texttt{TBabs1*cabs1}) and a second component for the clumpy 
partial coverer ($N_\text{H,2}$ = \texttt{TBabs2*cabs2}). The $N_\text{H,1}$ and $N_\text{H,2}$ 
column densities were left independent of each other. 
For spectral fitting the abundances of \citet{wilms:2000} and the cross-sections 
\citep{verner:1996} were used in all data sets.

Both observations show residuals to the continuum in the $35-40$\,keV energy range, which we interpret as 
the previously observed CRSF \citep{Mihara:1995, Kreykenbohm:2004}.  
Modeling these residuals with an absorption line 
with a Gaussian optical depth (\texttt{gabs}) improved the residuals significantly in 
both observations (see Figures~\ref{fig:spec010} and \ref{fig:spec020}). 
The best fit values of $E_\text{CRSF} = 42.5^{+3.2}_{-6.0}$\,keV and  $35.2^{+1.4}_{-0.9}$\,keV for Obs.~1 and Obs.~2, respectively, 
are consistent within 90\% confidence intervals. The widths of the CRSFs, 
$\sigma_\text{CRSF} = 9.5^{+1.6}_{-2.8}$\,keV (Obs.~1) and  $7.8^{+1.1}_{-0.9}$\,keV (Obs.~2), 
are also consistent within errors. An increase of $\sigma_\text{CRSF}$ with higher centroid energy, as previously observed in 
\textsl{RXTE} data \citep{Kreykenbohm:2004}, could not be detected, although the best fit values hint at such a behavior. 
Table~\ref{tab1} shows the best fit values for the continuum with a \texttt{gabs} CRSF line where the given errors are 90\% confidence values. 
To calculate the significance of the CRSF in the first and weaker observation, the null hypothesis approach was applied, 
where 10,000 spectra were created with Monte Carlo simulations using the best fit parameters without the CRSF. 
Gaussian uncertainties were used for the individual model parameter. 
Each model was fitted with and without the 
CRSF line to compare how much an inclusion of the line actually improves the fit. In $\sim 99.6\%$ of all fits, no CRSF feature were observed with a larger optical depth than the observed lower limits of $\tau = 10.3$, making the existence of the feature significant with $\sigma \sim 3$.
Including the CRSF in the real data, the best fit improved by a $\Delta \chi^2 = 100$.  A similar improvement of $\chi^2$ was only observed in $\sim 0.1\%$ ($\sigma \sim 3.3$) of the simulated spectra, concluding that the observed line in the first observation is indeed real. 
For the second observation, the line was even more pronounced, where the exclusion of the CRSF increased the $\chi^2$ by $\sim 200$.


\begin{table*}
\begin{center}
\caption{Phase averaged spectral parameters with the \texttt{GABS} model component for best fit spectrum\label{tab1}}
\begin{tabular}{lcccc} 
\hline
                      Parameter                                                             &  \multicolumn{2}{c}{403044010 (10ks)}                      	&   \multicolumn{2}{c}{403044020 (60ks)}   \\
                      & FI & BI & FI & BI\\
\hline
$N_\text{H,1} [10^{22}$ cm$^{-2}$]      		& $16.6^{+3.5}_{-3.3}$ 	& 16.6 (frozen)			& $20.9^{+3.5}_{-3.3}$		& 20.9 (frozen) \\
\text{Abund Ca}						     		& 1.55 (frozen) 			& 1.55 (frozen)			& $1.55^{+0.39}_{-0.18}$		& $1.98^{+0.41}_{-0.49}$ \\
\text{Abund Fe}							     		& 1.17 (frozen) 			& 1.17 (frozen)			& $1.17^{+0.04}_{-0.03}$		& $1.04^{+0.03}_{-0.04}$ \\
$N_\text{H,2} [10^{22}$ cm$^{-2}$]				 	& $76.9^{+3.9}_{-3.1}$ 	& 76.9 (frozen) 			& $28.4^{+1.0}_{-1.0}$		& 28.4 (frozen) \\ 
$\Gamma$	  								& $0.83^{+0.08}_{-0.06}$ 	&$0.82^{+0.09}_{-0.08}$	& $0.96^{+0.06}_{-0.04}$		& $0.98^{+0.03}_{-0.03}$\\ 
$A_\text{PL,1} $ $ [10^{-3}]^1$ fully covered			& $1.68^{+0.67}_{-0.50}$	&$1.38^{+0.20}_{-0.20}$	&$1.1^{+1.1}_{-0.5}$		& $4.70^{+0.91}_{-0.79}$\\ 
$A_\text{PL,2}$ $ [10^{-3}]^1$	part. covered 			& $92.5^{+16.1}_{-10.0}$ 	&$90.0^{+16.2}_{-8.1}$	& $215.5^{+25.1}_{-8.9}$	& $266.4^{+17.9}_{-15.5}$\\ 
$E_\text{cut}$ [keV] 			  					& 29.2 (frozen) 			& 29.2 (frozen) 			& $29.2^{+13.5}_{-2.1}$		& 29.2 (frozen) \\ 
$E_\text{fold}$ [keV]   							& $10.3^{+3.1}_{-3.0}$      & $8.3^{+2.0}_{-1.4}$ 	& $5.7^{+0.4}_{-2.5}$		& $5.6^{+0.2}_{-0.2}$\\ 
$E_\text{CRSF}$ [keV] 							& $42.5^{+3.2}_{-6.0}$     	& $39.7^{+3.7}_{-3.9}$ 	& $35.2^{+1.4}_{-0.9}$		& $34.9^{+0.9}_{-0.5}$\\ 
$\sigma_\text{CRSF}$ [keV]						& $9.5^{+1.6}_{-2.8}$     	& $8.6^{+2.0}_{-2.1}$	& $7.8^{+1.1}_{-0.9}$		& $7.4^{+0.5}_{-0.5}$ \\ 
$\tau_\text{CRSF}$ 							& $29.3^{+17.7}_{-19.0}$	& $18.9^{+14.8}_{-9.7}$ 	& $12.0^{+37.7}_{-4.5}$		& $10.6^{+1.8}_{-1.4}$\\ 
$E_{\text{S~K} \alpha}$ [keV]		 				& $2.33^{+0.01}_{-0.02}$ & $2.31^{+0.03}_{-0.03}$ 	& $2.33^{+0.01}_{-0.01}$		& $2.34^{+0.01}_{-0.01}$\\ 
$I_{\text{S~K} \alpha} [10^{-5}]^2$ 					& $2.34^{+0.61}_{-0.64}$ & $2.44^{+0.96}_{-0.89}$ 	& $5.36^{+0.50}_{-0.50}$		& $4.24^{+0.68}_{-0.67}$\\ 
EQW$_{\text{S~K} \alpha}^3	$ [eV]					& $860^{+1534}_{-585}$	& $1109^{+1313}_{-947}$&$671^{+491}_{-172}$		& $965^{+612}_{-445}$ \\
$E_{\text{Ar~K} \alpha}$ [keV] 					     	& $2.96^{+0.03}_{-0.03}$	& $3.02^{+0.06}_{-0.05}$	& $3.00^{+0.01}_{-0.	02}$		& $3.04^{+0.02}_{-0.02}$ \\ 
$I_{\text{Ar~K} \alpha} [10^{-5}]^2$ 					& $1.72^{+0.62}_{-0.69}$ & $1.00^{+0.94}_{-0.48}$ 	& $2.74^{+0.55}_{-0.54}$		& $4.53^{+0.94}_{-0.88}$\\ 
EQW$_{\text{Ar~K} \alpha}^3$ [eV]					& $179^{+502}_{-164}$	& $110^{+308}_{-110}$	&$38^{+29}_{-21}$			&$64^{+38}_{-37}$\\
$E_{\text{Ca~K} \alpha}$ [keV]		 				& $3.73^{+0.05}_{-0.04}$ & $3.76^{+0.07}_{-0.06}$ 	& $3.70^{+0.02}_{-0.01}$		& $3.71^{+0.03}_{-0.03}$ \\ 
$I_{\text{Ca~K} \alpha} [10^{-5}]^2$ 		 		& $1.71^{+0.77}_{-0.83}$ & $2.45^{+1.33}_{-1.35}$ 	& $4.72^{+0.95}_{-0.99}$		& $4.95^{+1.68}_{-1.71}$\\ 
EQW$_{\text{Ca~K} \alpha}^3$ [eV]					& $69^{+140}_{-69}$	&$100^{+154}_{-100}$	&$14^{+10}_{-8}$			&$14^{+14}_{-14}$\\
$E_{\text{Fe~K} \alpha}$ [keV]    			 & $6.399^{+0.007}_{-0.006}$& $6.410^{+0.010}_{-0.010}$& $6.409^{+0.003}_{-0.001}$&$6.424^{+0.003}_{-0.002}$ \\ 
$\sigma_{\text{Fe~K} \alpha}$ [eV]    	 			& $13^{+22}_{-2}$		& $<31$		& $<5$				&$<7$ \\ 
$I_{\text{Fe~K} \alpha} [10^{-4}]^2$			 		& $7.16^{+0.36}_{-0.22}$ & $6.96^{+0.63}_{-0.63}$ 	& $19.20^{+0.29}_{-0.30}$	& $22.9^{+0.61}_{-0.58}$\\ 
EQW$_{\text{Fe~K} \alpha}^3$ [eV]					& $241^{+53}_{-35}$	&$230^{+58}_{-52}$		&$133^{+27}_{-4}$			&$142^{+10}_{-10}$\\
$E_{\text{Fe~K} \beta}$ [keV] 						& $7.06^{+0.03}_{-0.03}$	& $7.04^{+0.04}_{-0.04}$	& $7.09^{+0.01}_{-0.01}$		& $7.09^{+0.02}_{-0.02}$ \\ 
$I_{\text{Fe~K} \beta} [10^{-4}]^2$ 					& $1.19^{+0.29}_{-0.35}$ & $1.54^{+0.59}_{-0.60}$ 	& $2.65^{+0.27}_{-0.25}$		& $4.50^{+0.53}_{-0.56}$\\ 
EQW$_{\text{Fe~K} \beta}	^3$ [eV]					& $43^{+33}_{-29}$		&$56^{+53}_{-51}$		&$21^{+7}_{-5}$			&$32^{+9}_{-10}$\\
$E_{\text{Ni~K} \alpha}$ [keV] 			   	  		& $-$ 		       		& $-$				& $7.46^{+0.02}_{-0.02}$		& $7.51^{+0.11}_{-0.06}$ \\ 
$I_{\text{Ni~K} \alpha} [10^{-4}]^2$ 			 		& $-$ 				& $-$			 	& $1.57^{+0.25}_{-0.25}$		& $0.67^{+0.62}_{-0.62}$\\ 
EQW$_{\text{Ni~K} \alpha}^3$ [eV]					& $-$				& $-$				&$15^{+9}_{-7}$			&$5.7^{+13.4}_{-5.7}$\\
Flux$_{2-10\,\text{keV}}^4$ absorb. & $1.61^{+0.04}_{-0.15}$&$1.63^{+0.03}_{-0.07}$	&$7.98^{+0.00}_{-1.42}$  		&$9.04^{+0.04}_{-0.06}$ \\
Flux$_{2-10\,\text{keV}}^4$ unabsorb. & $16.3^{+2.0}_{-1.1}$& $16.0^{+0.8}_{-0.7}$	&$31.8^{+1.3}_{-0.2}$  	&$35.9^{+1.0}_{-0.4}$ \\
$C_\text{XIS3}/C_\text{PIN}/C_\text{GSO} ^5$ 		& $0.93^{+0.02}_{-0.02} / 1.18^{+0.04}_{-0.04}$ /  -- &  --/ $1.12^{+0.07}_{-0.06}$ / --  & 
			$0.951^{+0.003}_{-0.003} / 1.31^{+0.02}_{-0.02} / 1.39^{+0.12}_{-0.12} $ 			& -- / $1.17^{+0.02}_{-0.02} / 1.27^{+0.11}_{-0.11} $ \\ 
$\chi^2$/dof     								& 260 / 245			& 145 / 130 			& 542 / 267   				& 328 / 150 \\ 
\hline
\multicolumn{5}{l}{(1)Units are ph keV$^{-1}$ cm$^{-2}$ s$^{-1}$, (2) Units are ph cm$^{-2}$ s$^{-1}$, (3) Values of EQWs are determined relative to the abs. continuum,} \\
\multicolumn{5}{l}{(4) Absorbed and unabsorbed flux units are  $[10^{-10}$ erg sec$^{-1}$ cm$^{-2}]$, (5) Values of $C$ are with respect to XIS 0 for the FI fits and XIS 1 for the BI fits. } 
\end{tabular}
\end{center}
\end{table*}

The CRSF can alternatively be described with the Lorenzian shaped 
\texttt{cyclabs} XSPEC model \citep{Mihara:1990}.  
\texttt{Cyclabs} is described by the centroid energy $E_\text{CRSF}$, the width $\sigma_\text{CRSF}$, and 
the resonance depth $\tau_\text{CRSF}$, similar to the \texttt{gabs} parameters.
The best fit continuum parameters are consistent with the
best fit values determined with the \texttt{gabs} component. The \texttt{cutoff} energy
with a best fit value of $E_\text{cut}=31.2^{+5.1}_{-2.9}$\,keV is consistent with the value obtained with the \texttt{gabs} model. 
The observed centroid energies of $E_\text{CRSF} = 35.0^{+3.0}_{-2.3}$\,keV and 
$31.6^{+1.0}_{-0.5}$\,keV are of the order of $10-20\%$ lower than the energies obtained with the \texttt{gabs} model. 
This discrepancy stems from a different calculation of the line centroid energy and is described in 
detail in \citet{Nakajima:2010}. The width $\sigma_\text{CRSF} = 11.2^{+3.8}_{-3.0}$\,keV 
and $12.3^{+1.7}_{-1.3}$\,keV for Obs.~1 and Obs.~2, respectively, is 
bigger than with the \texttt{gabs} model. 
The $\chi^2$ / dof
values of 260 / 245 and 548 / 267 for the two observations using the \texttt{cyclabs} model and 
thus were not a better fit when compared to the \texttt{gabs} model. 
For the final discussion we use the values determined by the \texttt{gabs} model. 

Several emission features were observed in the residuals of the fits in both observations and were subsequentially modeled with 
Gaussian emission lines (see Figures~\ref{fig:spec010} and \ref{fig:spec020}, inlay). 
The width of each line was set equal to that of the 
Fe K$\alpha$ emission line, while the intensities were left to vary 
independently. Energies were loosely constrained around the expected values of neutral material 
to avoid runaway of the line energies. 

A constant (\texttt{const}) was applied, taking small instrumental differences in the 
overall flux normalization into account. The constant was fixed at 1 for XIS 0 and was left 
free for the other instruments. $C_\text{XIS3}$, $C_\text{PIN}$ and $C_\text{GSO}$ in Table 1 are the 
cross calibration constants with respect to XIS\,0 for FI fits and with respect to XIS\,1 for the BI fits.

The final model had the form: 
\texttt{const}*\texttt{$N_\text{H,1}$} (\texttt{PL1}+ \texttt{$N_\text{H,2}$}*\texttt{PL2})* \texttt{fdcut}*\texttt{gabs}+ $6\times$\texttt{Gaussians}, 
where the power law indices of \texttt{PL1} and \texttt{PL2} are set to be equal to each other. The 
individual power law normalizations were independent and were used to calculate 
the fraction of the partial covering. 

Table \ref{tab1} summarizes the best fit values for both observations and for the 
individual FI / BI data sets. For the interpretation we will concentrate on the FI data for both 
observations, for reasons mentioned above. 

\subsection{Spectral results} 
The best fit spectral parameters were generally consistent with previous observations, with the exception of the cutoff energy. 
With a value of $29.2^{+13.5}_{-2.1}$\,keV, $E_\text{cut}$  was significantly higher than the 
$\sim 20$\,keV value measured in previous \textsl{RXTE} \citep{Mukherjee:2004} and \textsl{BeppoSAX} \citep{La-Barbera:2005}
observations, although these observations used slightly different spectral models for the cutoff energy. 
\citet{Kreykenbohm:2004} used also the here applied  Fermi--Dirac cutoff, resulting in cutoff energies of $10-15$\,keV. 
The folding energy for Obs.~1, $E_\text{fold} =10.3^{+3.1}_{-3.0}$\,keV, is consistent with values obtained for a similar 
orbital phase with \textsl{BeppoSAX} \citep{La-Barbera:2005}. In Obs.~2, the value of $E_\text{fold} =5.7^{+0.4}_{-2.5}$\,keV is consistent with the \textsl{RXTE} data for the pre-periastron flare and the periastron passage \citep{Kreykenbohm:2004}.

The observed column densities for the absorption due to the smooth stellar wind  ($N_\text{H,1}$) were consistent between both observations. 
A larger difference could be observed in the column density associated with the absorption due to the clumped wind ($N_\text{H,2}$), where 
the best fit value of Obs.~1 is almost a factor three larger than that of Obs.~2.  In addition to the H column density in the 
\texttt{TBnew} model, the relative Ca and Fe abundances were also left independent in Obs.~2. The best fit  values are measured from the
Ca and Fe K edges at $4.1$\,keV and $7.1$\,keV, respectively, and were slightly higher than solar abundances:  
$1.55^{+0.39}_{-0.18}$ for Ca and $1.17^{+0.04}_{-0.03}$ for Fe.
 For the first observation, the Ca and Fe 
abundances could not be well constrained and were fixed to 1.55 for Ca and 1.17 for Fe. 

The partial covering fractions can be calculated from the measured normalization values of the 
two power law components, $A_\text{PL1}$ and $A_\text{PL2}$: 
\begin{equation}
\text{Cvr. Frac.} = \frac{A_\text{PL2}}{A_\text{PL1} + A_\text{PL2}}. 
\end{equation} 
The factor is almost unity in both observations due to the dominance of the $A_\text{PL2}$ value. 
Similar large covering fractions have been observed in \textsl{XMM-Newton} data \citep{Furst:2011b} and 
in the \textsl{RXTE} data \citep{Mukherjee:2004}, indicating that the covering fraction does not change significantly 
for different parts of the orbit. 
The spectrum softened slightly from $\Gamma = 0.83^{+0.08}_{-0.06}$ for Obs.~1 to $0.96^{+0.06}_{-0.04}$ for Obs.~2.

The Fe K$\alpha$ line was detected at $6.409^{+0.003}_{-0.001}$\,keV (Obs.~2), which is, given the instrumental 
gain systematics, consistent with neutral Fe. In addition, the 
following lines have also been observed in both observations: S K$\alpha$ line, 
Ar K$\alpha$ line,
Ca K$\alpha$ line, Fe K$\beta$ line, and the Ni K$\alpha$ line, all with energies consistent with neutral 
material \citep[e.g.][]{Kaastra:1993}. Note that the Ni K$\alpha$ line was not detected in the fainter first observation. 
The observed line intensities are summarized in table~\ref{tab1} with 90\% confidence errors. 
The values are consistent between the FI and BI instruments. 
The widths of the lines were set to be equal to the Fe K$\alpha$ width within each observation, 
which was found to be $13^{+22}_{-2}$\,eV for Obs.~1 and had an upper limit of 5\,eV in Obs.~2 (FI values). 
The Compton shoulder to the Fe K$\alpha$ line, as observed by the \textsl{Chandra} 
observatory \citep{Watanabe:2003}, was not significantly detected in 
either observation. The observation of \citet{Watanabe:2003} was performed in the 
pre-periastron phase, where the luminosity was higher than 
in the two \textsl{Suzaku} observations. A Compton shoulder has also been 
observed with \textsl{XMM-Newton} in data taken during another
pre-periastron flare \citep{Furst:2011b}.

\section{Phase resolved analysis}
Barycentric and binary corrected light curves for different energy bands were extracted for both
observations using the orbital parameters from \citet{Koh:1997} with an updated orbital period and periastron time $t_0$ determined by \citet{Doroshenko:2010a}. 
Due to the strong variability of the pulse period on short time scales, individually determined pulse periods were used for 
each observation. An accurate pulse period could be established for 
Obs.~2,  but not for Obs.~1 due to its short duration of 10\,ks.
For the second observation the calculated 
pulse period had a value of $P = 685.4\pm0.9 $\,s. The  
\textsl{Fermi}/GBM instrument measures the pulse period of \gx on a regular basis, 
resulting in values of  $P = 687.28$\,s  for Obs.~1 and $P = 685.75$\,s  for Obs.~2 (Finger, 2011, priv. comm).
To create the pulse profiles, the  \textsl{Fermi}/GBM-provided pulse periods and epoch times for each observation 
were used.

\begin{figure}
 \plotone{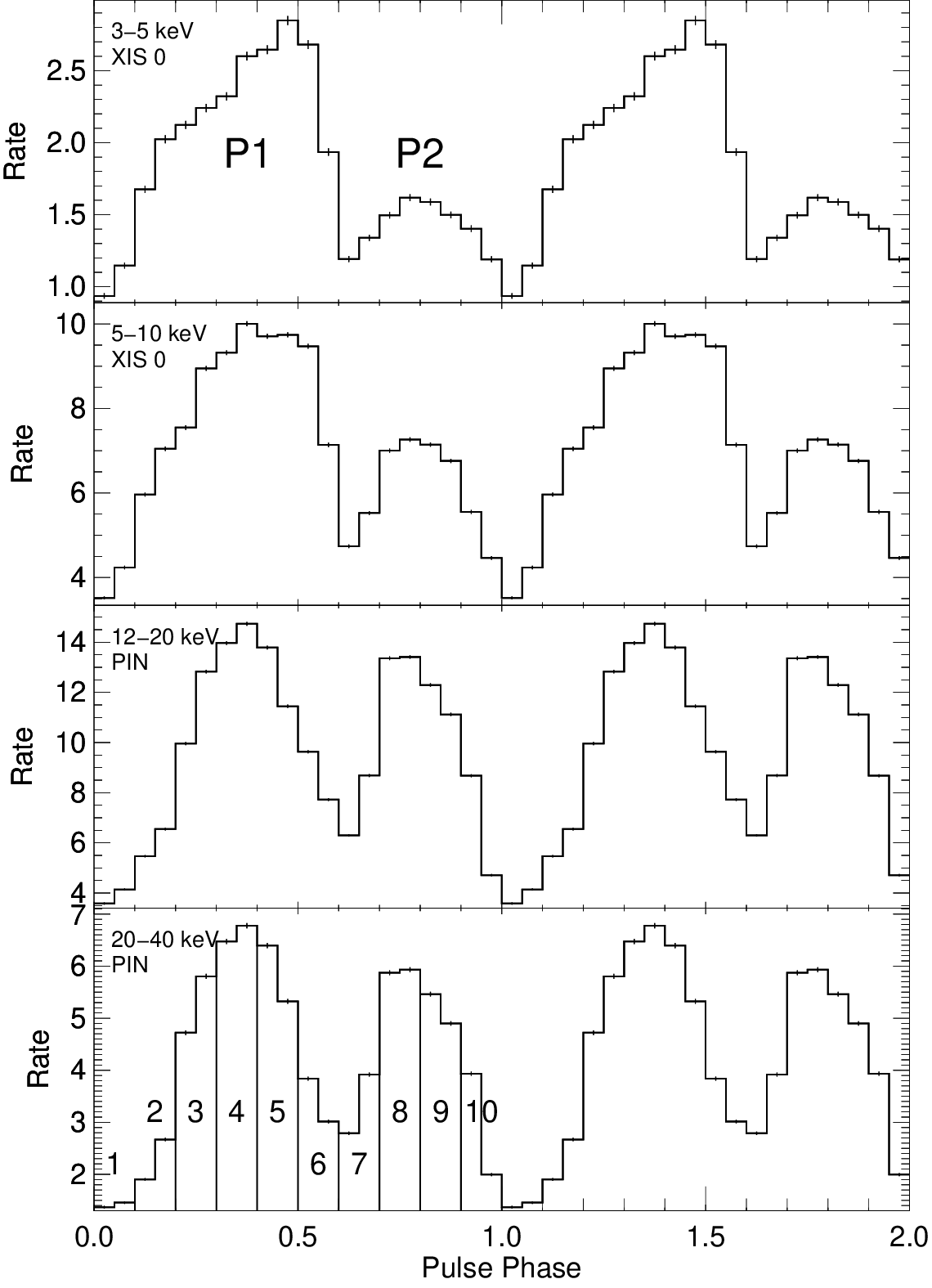}
\caption{Pulse profiles for different energy bands for Obs.~2. The two peaks are indicated in the top panel. The bottom pannel 
shows the selection of phase bins which were used for phase resolved spectroscopy.}
\label{fig:prof020}
\end{figure}

 We studied the energy dependent pulse profiles for Obs.~1 and Obs.~2. 
 Both pulse profiles are very similar, showing a double peak shape  
where the main peak (P1) is broader below 10\,keV. 
The second peak (P2) stays rather constant in width, but increases its relative intensity toward higher energies. 
As an example, the pulse profile for Obs.~2 in different 
energy bands is shown in  Figure~\ref{fig:prof020}. 
Comparing these pulse profiles with previous \textsl{RXTE} and \textsl{BeppoSAX} 
data showed that the general shape is consistent through all parts of the orbit.

\begin{figure}
 \plotone{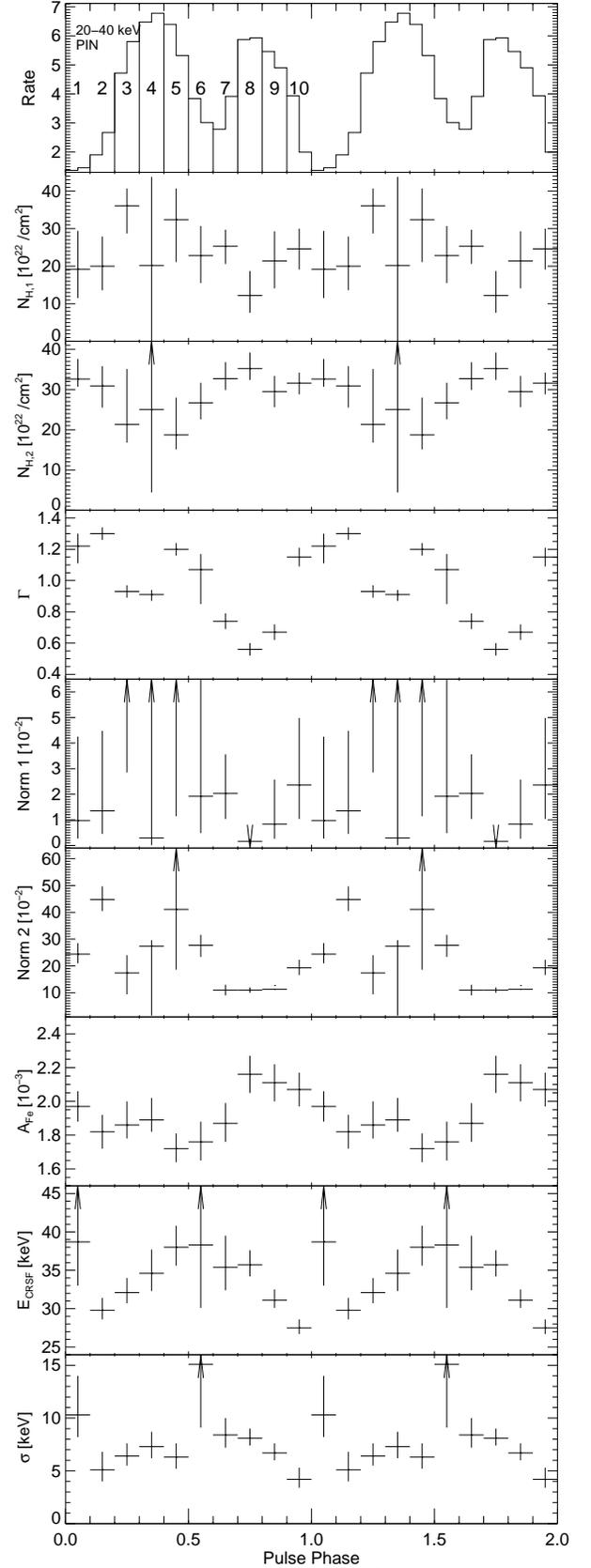}
\caption{Phase resolved spectral parameters for Obs.~2. Top panel shows the 
$20-40$\,keV pulse profile, where the individual phase bins are numbered. Units of power law normalization are ph 
keV$^{-1}$ cm$^{-2}$ s$^{-1}$. The intensity of the Fe line is units of ph cm$^{-2}$ s$^{-1}$. 
For clarity, two pulses are shown. }
\label{fig:phasespec1}
\end{figure}

The data of the longer Obs.~2 were divided into 10 equally spaced phase bins and individual spectra were extracted 
for the XIS and PIN instruments. Figure~\ref{fig:prof020} shows the individual phase bins in the lowest panel. 
This division resulted in an exposure time in each phase bin of $\sim 5-6$\,ks for each individual instrument. 
No GSO data were used in this analysis due to the reduced exposure time per phase bin. 
For spectral analysis, the phase averaged model was applied to 
all phase bins, resulting in best fit values summarized in Table~\ref{tab2}, 
as well as Figure~\ref{fig:phasespec1}. Again the cutoff energy was frozen to 
the phase averaged value of 29.2\,keV to avoid the previously discussed degeneracy 
with $E_\text{CRSF}$.

\begin{table*}
\begin{center}
\caption{Phase resolved spectral parameter from Obs. 2 for FI data. Same model as phase averaged data. \label{tab2}}
\begin{tabular}{lccc cccc ccc} 
\hline
Parameter   & PB1 &PB2&PB3& PB4&PB5&PB6&PB7 &PB8&PB9&PB10  \\
\hline
$N_\text{H,1} [10^{22}$/cm$^2$]      		&$19.2_{-7.7}^{+10.2}$	&$20.0_{-6.4}^{+7.9}$	&$36.1_{-7.4}^{+4.6}$	&$20.2_{-20.2}^{+23.6}$	&$32.4_{-11.3}^{+8.3}$	&$22.8_{-7.3}^{+7.9}$	&$25.3_{-4.7}^{+4.4}$	&$12.2_{-4.6}^{+6.5}$	&$21.4_{-7.3}^{+7.9}$	&$24.6_{-5.5}^{+5.4}$	\\ 

$N_\text{H,2} [10^{22}$/cm$^2$]		&$32.6_{-1.9}^{+5.0}$		&$30.9_{-5.4}^{+4.9}$	&$21.3_{-4.5}^{+13.8}$	&$25.0_{-20.6}^{+21.0}$	&$18.7_{-3.6}^{+9.3}$	&$26.7_{-4.1}^{+5.0}$ 		&$32.7_{-2.8}^{+4.1}$	&$35.2_{-2.8}^{+4.0}$	&$29.5_{-3.9}^{+3.9}$	&$31.6_{-2.8}^{+2.6}$	\\

$\Gamma$	  							&$1.22_{-0.11}^{+0.08}$	&$1.30_{-0.04}^{+0.04}$	&$0.93_{-0.04}^{+0.04}$	&$0.91_{-0.04}^{+0.03}$	&$1.20_{-0.04}^{+0.04}$	&$1.07_{-0.22}^{+0.10}$	&$0.74_{-0.05}^{+0.05}$	&$0.56_{-0.04}^{+0.04}$	&$0.67_{-0.05}^{+0.05}$	&$1.15_{-0.06}^{+0.06}$	\\

$A_\text{PL,1} [10^{-2}]^1$ ful. cov. 			&$0.97_{-0.70}^{+3.28}$	&$1.35_{-0.90}^{+3.13}$	&$10.12_{-7.27}^{+8.47}$&$0.29_{-0.28}^{+25.73}$&$8.69_{-7.55}^{+24.37}$&$1.92_{-1.44}^{+4.56}$	&$2.03_{-1.00}^{+1.53}$	&$0.16_{-0.27}^{+0.09}$	&$0.83_{-0.57}^{+1.74}$	&$2.36_{-1.33}^{+2.62}$	\\

$A_\text{PL,2} [10^{-2}]^1$ par. cov.			&$24.4_{-3.4}^{+4.1}$	&$44.8_{-4.3}^{+4.9}$	&$17.4_{-8.0}^{+6.6}$	&$27.4_{-25.9}^{+2.2}$	&$41.1_{-22.5}^{+77.1}$	&$27.7_{-4.3}^{+3.9}$	&$10.8_{-2.0}^{+2.0}$	&$10.6_{-1.0}^{+1.0}$	&$11.3_{1.4}^{+1.4}$	&$19.3_{-2.7}^{+3.0}$	\\

$E_\text{fold}$ [keV]   						&$7.4_{-2.3}^{+14.1}$		&$6.13_{-0.35}^{+0.38}$	&$5.6_{-0.3}^{+0.4}$	&$5.5_{-0.4}^{+0.5}$	&$5.9_{-0.6}^{+0.7}$	&$4.1_{-0.9}^{+5.0}$		&$4.5_{-0.6}^{+0.8}$	&$5.0_{-0.4}^{+0.5}$	&$4.4_{-0.3}^{+0.3}$	&$5.0_{-0.3}^{+0.4}$	\\

$E_\text{CRSF}$ [keV] 						&$38.7_{-5.7}^{+9.3}$		&$29.8_{-1.2}^{+1.6}$	&$32.1_{-1.4}^{+1.9}$	&$34.6_{-2.3}^{+3.1}$	&$38.0_{-2.4}^{+2.8}$	&$38.3_{-8.2}^{+22.9}$		&$35.4_{-3.0}^{+4.1}$	&$35.7_{-1.5}^{+1.9}$	&$31.1_{-1.0}^{+1.4}$	&$27.5_{-0.8}^{+1.1}$	\\

$\sigma_\text{CRSF}$ [keV]					&$10.3_{-2.1}^{+3.7}$		&$5.1_{-1.1}^{+1.7}$	&$6.4_{-0.9}^{+1.2}$	&$7.3_{-1.1}^{+1.4}$	&$6.3_{-1.1}^{+1.3}$	&$15.1_{-6.0}^{+15.0}$		&$8.4_{-1.2}^{+1.6}$	&$8.1_{-0.7}^{+0.9}$	&$6.7_{-0.7}^{+0.9}$	&$4.2_{-0.8}^{+1.1}$	\\

$\tau_\text{CRSF}$ 						&$33.4_{-16.7}^{+75.1}$	&$3.5_{-0.9}^{+1.5}$	&$6.7_{-1.5}^{+2.4}$	&$8.4_{-2.4}^{+4.7}$	&$8.3_{-3.4}^{+6.0}$	&$30.5_{-19.0}^{+30.0}$		&$17.0_{-5.2}^{+10.3}$	&$21.8_{-3.8}^{+6.0}$	&$11.4_{-1.9}^{+2.8}$	&$3.9_{-0.9}^{+1.2}$	\\

$I_{\text{Fe~K} \alpha} [10^{-3}]^2$ 			&$1.97_{-0.09}^{+0.09}$	&$1.82_{-0.10}^{+0.10}$	&$1.86_{-0.08}^{+0.14}$	&$1.89_{-0.07}^{+0.13}$	&$1.72_{-0.08}^{+0.09}$	&$1.76_{-0.11}^{+0.12}$	&$1.87_{-0.11}^{+0.12}$	&$2.16_{-0.11}^{+0.11}$	&$2.11_{-0.11}^{+0.11}$	&$2.07_{-0.10}^{+0.10}$	\\

Flux$_{2-10\,\text{keV}} ^3$& $4.93^{+0.03}_{-0.08}$&$7.77^{+0.08}_{-0.16}$	&$10.03^{+4.42}_{-0.61}$  		&$11.27^{+0.39}_{-3.37}$ & $10.93^{+4.37}_{-0.90}$ & $7.76^{+7.04}_{-3.52}$&$6.56^{+0.05}_{-0.62}$&$8.47^{+0.17}_{-0.20}$ & $7.62^{+0.10}_{-0.22}$ & $4.89^{+0.05}_{-0.16}$\\

C$_\text{XIS3}$/C$_\text{PIN}^4$ 				&0.94 / 1.15			& 0.94 / 1.15			& 0.95 / 1.19		&  0.94 / 1.21				& 0.94 / 1.21			& 0.94 / 1.19 					& 0.95 / 1.16			& 0.95 / 1.14			& 0.94 / 1.10			& 0.94 / 1.11			\\ 

$\chi^2$/dof     							&233 / 243				& 288 / 243			& 353 / 243			& 312 / 243			& 253 /243			& 303 / 243					& 283 / 243			& 292 / 243			& 264 / 243			& 247 / 243			\\ 

$\chi^2$/dof no CRSF 					&241 / 246				& 307 / 246			& 391 / 246			& 341 / 246			& 273 / 246			& 305 / 246					& 291 / 246			& 344 / 246			& 297 / 246			& 270 / 246			\\

\hline
\multicolumn{10}{l}{(1) Units are ph keV$^{-1}$ cm$^{-2}$ s$^{-1}$, (2) Units are ph cm$^{-2}$ s$^{-1}$, (3) Units are $10^{-10}$ erg cm$^{-2}$ sec$^{-1}$ , (4) Values of $C$ are with respect to XIS 0 for the FI fits.}

\end{tabular}
\end{center}
\end{table*}

The two absorbing components, $N_\text{H,1}$ from the smooth stellar wind and 
$N_\text{H,2}$ from the partial coverer vary between 20 and 40$\times 10^{22}$ cm$^{-2}$, 
but show an anti-correlated trend throughout the 
first peak. Although the error bars are rather big, one can see an indication that 
$N_\text{H,1}$ follows the flux in P1, whereas the $N_\text{H,2}$ value dips 
at the same time.   

In contrast to \citet{Kreykenbohm:2004}, the power law index $\Gamma$ is varying 
strongly throughout the pulse. The values are 0.9--1.3 throughout the first peak and 
drop suddenly to 0.6--0.8 for the second peak (Figure~\ref{fig:phasespec1}).  
Measuring the second peak to be significantly harder than the first peak supports the 
behavior seen in the pulse profile, where the intensity of P2 increases at higher 
energies, becoming similar to the intensity in P1. 
Power law normalizations are relatively small and badly constrained for the first 
power law component. The partially covered power law normalization seems to 
follow the flux, showing a higher value throughout the first peak. 
The folding energy does not vary significantly throughout the orbit and shows values 
between 5 and 6\,keV (not shown in Figure 7)

The CRSF energy varies between 30--40\,keV, similar to the values observed 
in the \textsl{RXTE} data \citep{Kreykenbohm:2004}. 
Table 2 shows that the addition of the CRSF in the individual spectra improved 
the $\chi^2$ for most of the phase bins, except in phase bin 6,
which falls in the gap between both P1 and P2. Phase bins 1 and 7, also minima 
in the pulse profile, only show a small improvement in the best fit when the CRSF 
was included. The CRSF energy is not very well constrained in phase bins 1 and 
6 due to the lack of sufficient statistics. For the geometrical discussion below, 
these two phase bins are ignored.
In all other phase bins the addition of the CRSF improves the fit. 
The energy changes very smoothly throughout the 
pulse and does not follow explicitly the observed flux in each phase bin. The 
CRSF energy increases during the main peak and reaches the maximum at the 
falling flank of P1. During P2 the CRSF energy decreases until it reaches a 
minimum at the dip between P2 and P1. 

To estimate the significance of the CRSF in the phase resolved analysis, again 
the null hypothesis method was applied in two out of the 8 remaining phase bins, 
which are good representatives of all data points. Phase bin 2 is a good example 
for a shallow CRSF, whereas phase bin 5 is an example for a phase bin with a 
higher flux. Similarly as with the phase averaged data, 10,000 spectra 
were simulated and modeled without and with the CRSF. The interpretation of 
the observed feature as being due to stochastic fluctuations  with a depth similar 
to the observed values could be ruled out with a probability of over $99\%$ 
($\sigma \sim 3$) in both phase bins. This leads to the conclusion that the CRSF 
can be observed with sufficient significance to allow the interpretation below.

\section{Discussion}

\subsection{Phase averaged continuum } 
In both observations very strong absorption can be observed. $N_\text{H,1}$, which is the absorption due to the 
smooth stellar wind, is consistent within both observations. $N_\text{H,2}$, the absorption from the partial coverer, 
is significantly stronger in the first observation, establishing the existence of an additional component in the line of sight. 
\citet{Leahy:2008} used \textsl{RXTE}/ASM and PCA data to show that a stellar wind and a stream model component  can describe the 
observed count rate throughout the orbit and that the increase of the column density at orbital phase $\sim 0.2$ is mainly 
due to the stream component. 
The findings in both \textsl{Suzaku} observations are supportive of this theoretical picture. 

\gx shows very strong luminosity variability on very short time scales \citep{Kreykenbohm:2004, Furst:2011b}, where the 
column density increases to up to 50\% in periods of lower activity. Both \textsl{RXTE} and \textsl{XMM-Newton} data showed 
a much higher column density in the pre-periastron flare than in the two data points observed with \textsl{Suzaku}. 
One possible explanation is that the wind is indeed so variable and clumpy that these variations are just not properly predictable, especially 
in times of high activity, such as the pre-periastron flare. \citet{Leahy:2008} investigated \textsl{RXTE}/PCA $N_\text{H}$ values of the absorbed component from 
archival observations for different parts of the orbit and found no increased column density in the  pre-periastron flare. 
Observations throughout one full binary orbit could help to understand the variations in $N_\text{H}$ on time scales of days.  

The observed luminosity dependence of the power law index $\Gamma$ and the folding 
energy is similar to other sources, such as V0332+53 \citep{Mowlavi:2006} and 4U0115+63 
\citep{Tsygankov:2007}. The power law index $\Gamma$ shows a hardening for 
the first observation, whereas the folding energy $E_\text{fold}$ is slightly higher 
when compared to Obs.~2. \cite{Soong:1990} observed the $E_\text{fold}$ variation 
in Her X-1 phase resolved  \textsl{HEAO-1} data and concluded that the parameter is 
dependent on the viewing angle of the accretion column and can be used to describe 
the plasma temperature of the system. In the case of \gx the smaller $E_\text{fold}$ value 
in Obs.~2 could indicate a lower plasma temperature which can be interpreted that the 
X-ray emission region is further up in the accretion column \citep{Basko:1976} (see also Section 5.2). 

The softening of the power law index $\Gamma$ with increased luminosity is also in agreement with the basic model of 
the accretion column that the plasma temperature decreases with increased height. With increased luminosity, the 
rate of accreted material would increase and the amount of soft photons created by the lateral walls of the 
relatively taller column would increase, leading to a softer spectrum, as is observed. 

For the  second observation the abundances for Fe and Ca were left independent and a slight overabundance was observed. 
Taking into consideration that the abundances used in \citet{wilms:2000} are derived from 
the interstellar medium (ISM) a small overabundance from an evolved star may be expected.

\subsection{Variations in the CRSF parameters}
In both observations the CRSF was clearly observed and the \texttt{gabs} absorption component 
improved the overall fit significantly. 
Strong magnetic fields ($B \sim 10^{12}$\,G) exist close to the 
NS magnetic poles, where photons 
at energies close to the Landau levels are resonantly scattered from the line of sight
and result in an absorption line-like feature in the spectrum. This feature 
provides a direct method to measure the magnetic field strength close to the 
NS surface, where the fundamental centroid energy can be described as 
\begin{equation} 
E_\text{CRSF} = 11.6\text{\,keV} \times \frac{1}{1+z} \times \frac{B}{10^{12}\text{\,G}}
\end{equation} 
where $z$ is the gravitational redshift at the scattering site, with $z \approx 0.3$ for typical NS.  
Using the $E_\text{CRSF}$ values obtained in the phase averaged spectra, the calculated magnetic field 
has a strength of $4.76^{+0.36}_{-0.67} \times 10^{12}$\,G (Obs.~1) and $4.10^{+0.15}_{-0.10}\times 10^{12}$\,G (Obs.~2) which 
is consistent within errors for both observations. 

The CRSF parameters are consistent with previous results using \textsl{RXTE} data and are lower than 
the observed \textsl{BeppoSAX} \citep{La-Barbera:2005} and \textsl{INTEGRAL} \citep{Doroshenko:2010} 
values of $45-50$\,keV. These observations were obtained during pre-periastron outburst,
where the luminosity of the source is much higher than in the two \textsl{Suzaku} observations presented in this paper. 
A luminosity dependence of the phase-averaged CRSF centroid energy 
is observed in multiple other sources, where an anti-correlation between CRSF and luminosity was observed in V\,0332+53 and 4U\,0115+63
\citep{Tsygankov:2006,Mihara:2007}, and  a positive correlation was observed in Her~X-1 \citep{Staubert:2007}. 
\citet{Klochkov:2011} have confirmed, from pulse to pulse variability studies, such correlations for V\,0332+53, 4U\,0115+63 and Her~X-1, and also found an anti-correlation for A\,0535+26. 
Although the CRSF values are consistent with a single value in both \textsl{Suzaku} observations, 
a hint of an anti-correlation can be seen in the data. 
Compared with the \textsl{BeppoSAX} and \textsl{INTEGRAL} data, the CRSF centroid energy 
might have a positive correlation. 
Parallel to the CRSF-luminosity correlation, \citet{Klochkov:2011} also observed in pulse to pulse data of four sources that the 
power law index $\Gamma$ shows an opposite luminosity 
dependence than the $E_\text{CRSF}$. In the \textsl{Suzaku} observations discussed here, 
$\Gamma$ is observed to soften with increased luminosity, another indication that a possible negative CRSF-luminosity correlation exists. 

The two opposite scenarios seem to depend on the 
luminosity, i.e., if the observed luminosity is below or above a critical luminosity (CL) which can be derived from 
the Eddington luminosity of a system \citep{Becker:2007}. The CL is of order  $\sim 10^{37}$ erg s$^{-1}$, but depends 
also on the accretion geometry (spherical or disc) as well as the height and diameter of the accretion column. 

Above the CL, the infalling proton density becomes so large that the protons 
begin to interact and decelerate, creating a radiation pressure dominated 
shock region above the magnetic pole. This region of increased density is 
most likely the region where the CRSF is created \citep{Basko:1973}. 
With increasing luminosity, the shock region moves higher up 
in the accretion column, where a smaller local magnetic field value results in a 
lower observed CRSF centroid energy, as is observed in V\,0332+53 or 4U\,0115+63. 
The observed CRSF-luminosity dependence as well as the $\Gamma$-luminosity 
dependence in \gx is very similar to the scenario described here. 

Below the CL, the accreting matter slows down via hydrodynamical shock, `Coulomb friction' 
or nuclear collision \citep{Basko:1973, Braun:1984}. 
With increasing accretion rate, the luminosity increases and the deceleration region is pushed 
closer to the NS surface, where the magnetic fields are higher. This would result in a 
positive CRSF-luminosity correlation, as observed in Her~X-1 and A\,0535+26. 

With a distance of 3\,kpc, the intrinsic unabsorbed $2-10$\,keV luminosity of $\sim 2\times10^{35}$\,erg\,s$^{-1}$
is significantly below the typical CL of $\sim 10^{37}$\,erg\,s$^{-1}$. Although it has been observed that 
sources with similar luminosities can show the opposite correlation \citep{Klochkov:2011} due to 
individual critical luminosity values, the large luminosity difference compared to CL puts this source 
in the same regime as Her~X-1 and A\,0535+26 and at odds with the notion of a well-defined CL. 
The observed variation in $\Gamma$ and possibly in $E_\text{CRSF}$ 
would then stem from another mechanism. 

\subsection{Emission lines} 
The existence of multiple fluorescence emission lines, especially at lower energies, where the absorption is very 
dominant, indicates that the source of the emission originates from a region where the column density is not 
very large, i.e. the outer layers of the stellar wind in our line of sight \citep{Furst:2011b}. If the emission lines would be embedded deeper in 
the stellar wind, the lines, especially at lower energies would have to be significantly stronger, to 
be detected at all. On the other hand, if the line emitting region is in the 
outer layers of the wind, the incident soft X-ray flux would 
be drastically reduced, making the equivalent width very large, as observed 
in the low energy emission lines (Table~1). A possible explanation would be that the emission  
region is very large for the lines, and maybe spread over the entire surface of the stellar wind. 

The most dominant line emission stems from the Fe~K$\alpha$ transition at $6.4$\,keV. 
The observed energies in both observations are consistent with the 
emission in neutral material. The ratio of the intensities of Fe~K$\beta$/Fe~K$\alpha$ for both 
observations, $0.16^{+0.05}_{-0.04}$ for Obs.~1 and $0.14^{+0.01}_{-0.01}$ for Obs.~2, is consistent with the fluorescing
material being neutral or only slightly ionized \citep{Kaastra:1993}. 
The equivalent widths (EQW) of the two Fe K$\alpha$ lines show that the intensity relative to the continuum 
is reduced by a factor of $\sim2$ in the second observation. 
Based on emission line widths, \citet{Endo:2002} estimated in \textsl{ASCA} data that the Fe emission  
originates within $10^{10}$\,cm of the continuum emission source. 
\cite{Furst:2011b} use \textsl{XMM-Newton} pulse by pulse observations to observe a correlation with 
increased flux of the source, showing that the Fe emission region is not far from the X-ray source. 
In the \textsl{Suzaku} observation, however, the Fe K$\alpha$ line flux does not directly follow the observed luminosity. 
The increase in absorbed $2-10$\,keV luminosity is a factor of $\sim 5$, whereas the increase in the fluorescence 
line intensities is only $2-3$. This difference in intensity change indicates that the distance between continuum and 
fluorescence emission lines is large, especially when compared to the proposed $\sim0.3$ lt-s from \citet{Endo:2002}. 
In the 6 months' time between the two observations, however, the overall emission geometry could have changed, 
accounting for the difference in the fluorescence intensities. If the emission originates from a 
confined region in the stellar wind, observations taken during two different 
orbital phases would each yield a different distance to this region.

In addition to the strong Fe K$\alpha$ line, many other emission lines are observed 
in the XIS spectra in both observations. The observed intensities of these lines show 
the same behavior, where the flux for the second observation is $\sim 50\%$ or less compared to the 
first observation, although the observed uncertainties for the different lines, especially for 
S~K$\alpha$ and Ar~K$\alpha$, are very large. Furthermore, the S~K$\alpha$ line falls into an energy regime 
where calibration uncertainties have been previously observed \citep[e.g.,][]{Suchy:2011}, so that a 
detailed analysis is not possible at this time.

\subsection{Pulse profiles} 
Historically, \gx shows very strong pulse to pulse variations and an intensive flaring behavior on very short time 
scales \citep{Kreykenbohm:2004, Furst:2011b}, e.g., throughout the pre-periastron flare. 
The observations obtained here do not show such flaring behavior,
and only show the regular pulsation throughout the observation. 
A pulse period could only be established for Obs.~2 of the \textsl{Suzaku} data, which was consistent with \textsl{Fermi}/GBM data. 
With the GBM pulse period for the first observation, pulse profiles for both observations could be 
produced for different energy bands. 
The two peaked pulse profiles show a similar behavior for both observations, where the second peak 
is getting strong at higher energies. This behavior is in contrast to many similar sources, such as 
4U~0115+63 \citep{Tsygankov:2007}, V~0332+53 \citep{Tsygankov:2006}, 4U\,1909+07 \citep{Furst:2011}, and 1A1118$-$61 \citep{Suchy:2011}, 
where the pulse profile turns into a single peak profile at higher energies.  A reduction in the intensity of 
the second pulse has also been observed in \textsl{BeppoSAX} data for different parts of the orbit \citep{La-Barbera:2005}.

\subsection{Geometrical constraints using a simple dipole model} 
The phase resolved analysis of the second observation showed strong variations in several spectral parameters 
throughout the pulse profile. As previously observed by \citet{Kreykenbohm:2004}, the CRSF centroid energy 
varies by $\sim 30\%$ where the highest energy is detected at the peak and the 
falling flank of P1. Such behavior has been observed in multiple other 
sources, e.g., Vela~X-1 \citep{La-barbera:2003}, 4U~0352+309 \citep{Coburn:2001}, 
and Her~X-1 \citep{Klochkov:2008}. A model for the CRSF variations is based upon  
the change in the viewing angle throughout the pulse and thus  
different heights of the accretion column are probed, yielding a different local 
magnetic field observed for each phase bin. 

A simple approach to derive a possible geometry for the neutron star and the magnetic field uses 
the variation in the observed magnetic field throughout the pulse phase. 
In this case the very smooth and sinusoidal variation (Figure~8) can be modeled as a simple dipole 
where the total magnetic moment $\mu$ is calculated from the phase averaged CRSF energy value. 

\begin{figure}
 \plotone{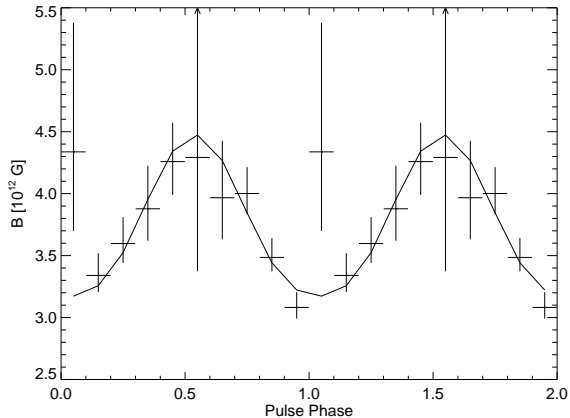}
\caption{Calculated phase resolved magnetic fields from the CRSF energy and the best fit 
for a simple magnetic dipole model.}
\label{fig:magnet}
\end{figure}

The variation of the CRSF energy is then fitted by changing the geometrical angles of the system until a best fit converges.
Appendix A discusses the model in detail, introducing the 
three free parameters: $\Theta$, the viewing angle between line of sight and the NS spin axis, 
the inclination angle $\alpha$ of the magnetic moment with respect to the spin axis, and the angle $\beta$,
indicating the 'lag' of the $B$-field plane with respect to the ephemeris, i.e., 
the observed shift in pulse phase (see also Figure~9)
The best fit values for $\beta$ were either $-4.6^\circ$ or $175.4^\circ$, depending on the starting values of the fit (see Appendix A), 
 where the latter value is equivalent to the former when the dipole polarity is flipped.
$\alpha$ and $\Theta$ angles showed a very strong interdependence, and 
the best fit $\Theta/ \alpha$ pairs were:  $-14.6^\circ / 67^\circ$, $14.6^\circ/113^\circ$, 
$67^\circ/-14.6^\circ$, and $113^\circ/14.6^\circ$.  
All best fit values have a $\chi^2 \approx 4$ with 7 dof. 
These pairs do show a degeneracy of $0\pm14.6^\circ$ and $90\pm23^\circ$, which can be explained 
with a geometrical symmetry,  when calculating the magnetic field for each individual phase bin.
The first and last two pairs can be each treated as the same geometry, rotated by 180 degrees. 
 Figure~\ref{fig:magnet}
shows the calculated magnetic field values from the CRSF centroid energies 
and the theoretical values for $B$ from the best fit angles for a simple dipole model with 
$\Theta = 67^\circ$, $\alpha = -14.6^\circ$ and $\beta = -4.6^\circ$.

The phase resolved energy and width of the CRSF show a strong correlation, 
where the width varies in phase with the CRSF energy (Figure.~7). \citet{Kreykenbohm:2004} 
observed a similar correlation,
where the magnetic pole was observed under different viewing angles $\Phi$, where $\Phi$ is the angle 
between the line of sight and the magnetic axis. 
\citet{Meszaros:1985} showed that the anisotropic velocity field of the electrons 
in the accretion column leads to a fractional line width of:
\begin{equation}
\frac{\sigma_\text{CRSF}}{E_\text{CRSF}}\approx \sqrt{8 \text{ln}2 \frac{kT_\text{e}} {m_\text{e}c^2}}  |\text{cos} \Phi|=k |\text{cos}\Phi |.
\end{equation}

The $\sigma_\text{CRSF}/E_\text{CRSF}$ ratios vary in the 0.16 -- 0.24 range, 
corresponding to variations of $\pm 20\%$ throughout the pulse phase. The two 
outliers with large errors in phase bins 1 and 6 were ignored. 
The variation of $\Phi$ can be estimated as $\Theta-\alpha$ and $\Theta+\alpha$ throughout one 
pulse phase, where the magnetic pole rotates around the NS spin axis, 
tilted by $\Theta$ with respect to the line of sight. 

Using the average ratio of 0.2, we calculated the values of $k$ for the 
two geometries with $\Theta =-14.6^\circ$ ($k = 0.21$) and $\Theta = 67^\circ$ ($k = 0.51$). 
For each geometry, we used the variation of the $\sigma_\text{CRSF}/E_\text{CRSF}$ ratio ($\pm 0.04$)
to calculate the variation in $\Phi$, resulting in values of $\Phi = 15^\circ\pm25^\circ$ and 
$\Phi = 67^\circ\pm5^\circ$. These variations are smaller then the values expected from the 
the geometrical discussion, but show a similar behavior, where the variation of the angle 
is larger for the smaller value of $ \Phi$. 
From Equation 4, we were also able to estimate a possible plasma temperature for the 
two geometries from the calculated values of $k$.
The plasma temperature of $kT \sim 4$\,keV for $\Theta = 15^\circ$ 
is very similar to the observed folding energy $E_\text{fold}$, which is an indication 
of the plasma temperature \citep[][and references therein]{Burderi:2000}. 
The value of $kT \sim 24$\,keV for the $\Theta = 67^\circ$ geometry is much larger and 
is not consistent with the observed $E_\text{fold}$. 
Assuming that the NS spin axis is aligned with the inclination of the binary system, a value of $\Theta = 15^\circ$ would 
put the inclination at $i \sim 75^\circ$, a value which is only marginally above the upper limit of $72^\circ$, as determined 
by \citet{Kaper:2006}. 

The strong Fe line was detected in all 10 phase bins. The measured line flux did 
not change significantly throughout the pulse phase, while the $2-10$\,keV flux varied by more than a factor of 2. 
From this we conclude that the distance to the Fe fluorescence region is greater than $\sim 700$ lt-s 
($\sim 2 \times 10^{13}$\,cm), from the NS. This is in conflict with the conclusion of \citet{Endo:2002} 
that the emission region must be closer than $10^{10}$\,cm from the NS, based upon the width 
of the emission lines measured by ASCA. Their assumption is that the emission region is close to the NS 
and that the material free falls onto the NS surface, encountering much faster velocities, which caused the 
observed broadening of the lines. 
The smaller measured Fe K$\alpha$ line width in the phase averaged spectra is more 
consistent with the assumption that the line broadening is due to the terminal velocity of $300-400$\,km 
of the line driven wind \citep{Parkes:1980}.

\section{Summary} 
We presented results of two \textsl{Suzaku} observations of \gx taken shortly after the periastron passage. 
The spectra were modeled with a partially and fully covered power law with a Fermi-Dirac cutoff and a CRSF at $\sim 35$\,keV. 
The column density of the smooth stellar component only marginally changed and the clumpy wind component significantly changed between both observations. 
Flux dependencies of the CRSF and $\Gamma$ were not significant, but do hint at similar correlations such as those observed in V\,0332+53 and 4U\,0115+63, although the observed luminosity is significantly below the calculated critical luminosity of this system. They are  in the realm where one might expect the energy of the CRSF to decrease with decreasing flux.  
The variations in the pulse profiles and the CRSF throughout the pulse phase have a signature of a dipole magnetic field. 
Using a simple dipole model we calculated the expected magnetic field values for different pulse phases and 
were able to extract a set of geometrical angles, loosely constraining the dipole geometry in the NS. 
Model constraints derived using the observed ratio of $\sigma_\text{CRSF}$ to $E_\text{CRSF}$, together with calculated plasma temperatures, 
favor a solution wherein the spin axis is tilted $\sim 15^\circ$ to our line of sight.

\begin{acknowledgements}
SS acknowledges support by NASA contract NAS5-30720 and grant NNX08AX83G. 
FF is supported by DLR grant 50 OR 0808 and via a DAAD Fellowship.
IC acknowledges financial support from the French Space Agency CNES through CNRS. 
This research has made use of data obtained from the Suzaku satellite, a collaborative mission 
between the space agencies of Japan (JAXA) and the USA (NASA).
\end{acknowledgements}

\appendix

\section{A. Geometrical dipole model}
The variability of the CRSF energy throughout the pulse phase can be used to understand the geometry of the 
neutron star. We assume a simple dipole magnetic field around a spherical neutron star, however
we want to emphasize that these assumptions are very rudimentary, and effects such as 
the displacement of the dipole field from the neutron star center are not taken into account here.

The total magnetic moment $\mu$ of the neutron star can be calculated with: 
\begin{equation}
\langle \mu \rangle = B \cdot R_\text{NS}^3 \langle \sqrt{1+3\text{cos}^2\Phi} \rangle= 6.776\times10^{30}\,\text{G cm}^3
\end{equation}
using the average magnetic field from the phase averaged CRSF centroid energy ($B=4.1 \times10^{12}$\,G)
and a typical neutron star radius of $R_\text{NS} = 10$\,km. The value of the last part, including the azimuthal angle $\Phi$ of 
the magnetic pole was estimated to the average value of 1.54.

\begin{figure}
 \plotone{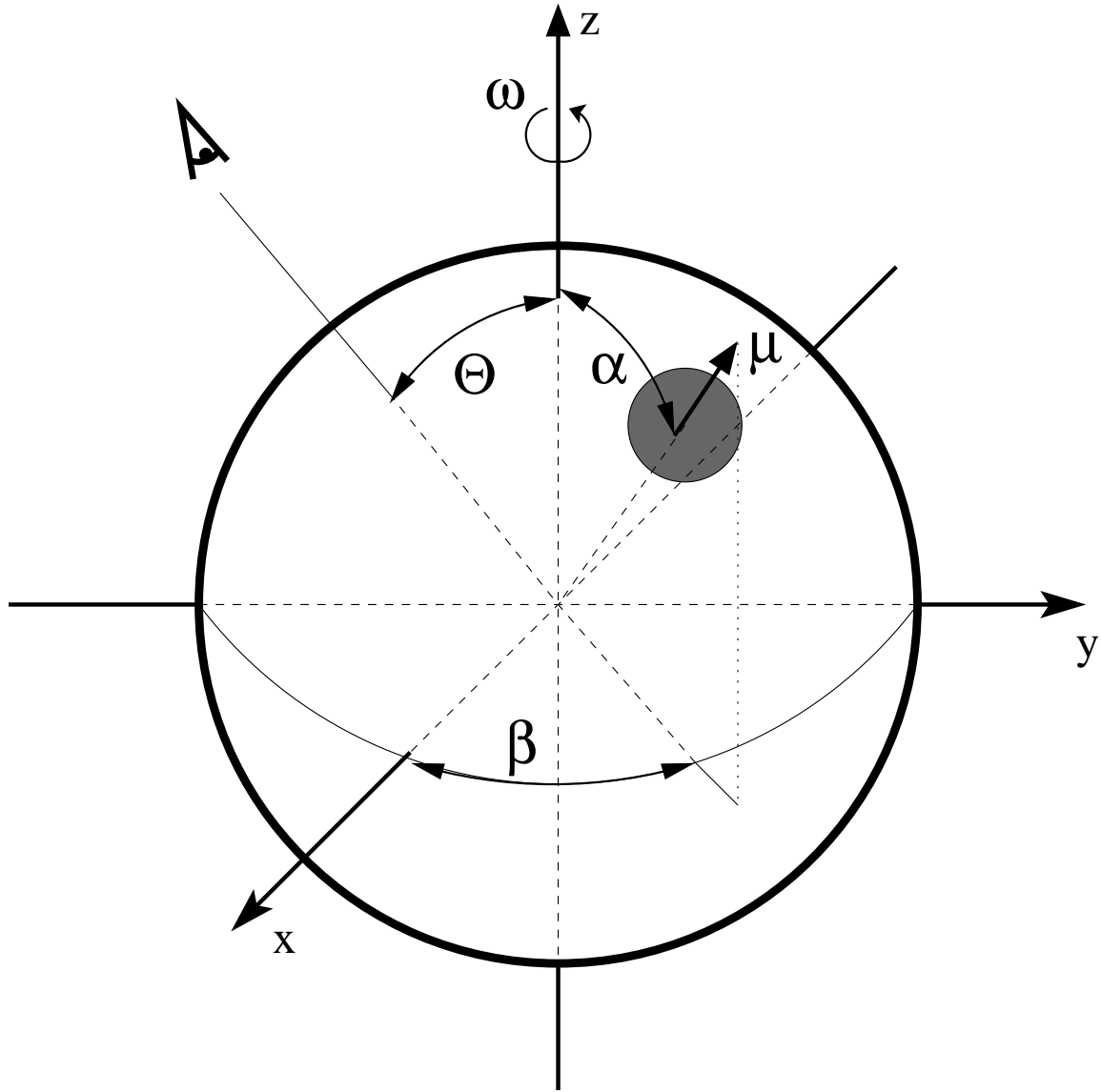}
\caption{Schematic view of the dipole model, indicating the different angles used to calculate the CRSF energy variation.$\beta$ is the angle between the projection of the magnetic moment in the x-y-plane and the ephemeris (x-axis) .The eye indicates the line of sight towards us.}
\label{fig:nsmodel}
\end{figure}

The total average magnetic moment is divided into three components based on the 
cartesian coordinate system in 
Figure~\ref{fig:nsmodel}: 
\begin{equation}
\bold{m} = \mu \begin{pmatrix} \text{sin}(\alpha) \text{cos}(\beta) \\  \text{sin}(\alpha) \text{sin}(\beta) \\ \text{sin}(\beta) \end{pmatrix}\end{equation}
where $\alpha$ is the angle between the dipole field and the spin axis and $\beta$ is the rotation 
angle from the ephemeris, i.e. the x-axis. 
Next, the vector $\bold{n}$ was calculated for each 
individual phase bin $\phi$ used in the phase resolved analysis. 
\begin{equation}
\bold{n} =  \begin{pmatrix} \text{sin}(\Theta) \text{cos}(\phi) \\  \text{sin}(\Theta) \text{sin}(\phi) \\ \text{sin}(\phi) \end{pmatrix}
\end{equation}
The individual components of $\bold{n}$ are the projections of each phase bin on the three axes, as seen from the line of sight
of the observer. 
The angle $\Theta$ corresponds to the angle 
between the line of sight and the spin axis.
For each of the individual phase bins, the magnetic field  was calculated using the position vector $\bold{n}$ indicating the 
position of the magnetic pole and the vector $\bold{m}$, which indicates the magnetic moment:
\begin{equation}
B_\text{phase} =  \frac{3\cdot \bold{n}\left|\bold{m}\cdot\bold{n}\right|}{\left|\bold{n}\right|^5}-\frac{\bold{m}}{\left|\bold{n}\right|^3}
\end{equation}

The calculated magnetic field values were then fitted to the measured magnetic field, varying the 
three angles $\Theta$, $\alpha$, and $\beta$ until the best fit converged.
To avoid wrong solutions due to local minima, the starting parameter of all three angles were 
varied in steps of  $10^\circ$ before the fit was started. 
The best fit solutions resulted in the 
values that are discussed in detail in section 5.5.


\begin{thebibliography}{}

\bibitem[\protect\astroncite{{Basko} \& {Sunyaev}}{1973}]{Basko:1973}
{Basko}, M.~M., \& {Sunyaev}, R.~A.,  1973, \apss, 23, 117

\bibitem[\protect\astroncite{{Basko} \& {Sunyaev}}{1976}]{Basko:1976}
{Basko}, M.~M., \& {Sunyaev}, R.~A.,  1976, \mnras, 175, 395

\bibitem[\protect\astroncite{{Becker} \& {Wolff}}{2007}]{Becker:2007}
{Becker}, P.~A., \& {Wolff}, M.~T.,  2007, Astrophys. J., 654, 435

\bibitem[\protect\astroncite{{Braun} \& {Yahel}}{1984}]{Braun:1984}
{Braun}, A., \& {Yahel}, R.~Z.,  1984, \apj, 278, 349

\bibitem[\protect\astroncite{{Burderi} et~al.}{2000}]{Burderi:2000}
{Burderi}, L., {Di Salvo}, T., {Robba}, N.~R., {La Barbera}, A., \&
  {Guainazzi}, M.,  2000, \apj, 530, 429

\bibitem[\protect\astroncite{{Coburn}}{2001}]{Coburn:2001}
{Coburn}, W.,  2001,
\newblock Ph.D. thesis, UC San Diego

\bibitem[\protect\astroncite{{Doroshenko} et~al.}{2010a}]{Doroshenko:2010a}
{Doroshenko}, V., {Santangelo}, A., {Suleimanov}, V., {Kreykenbohm}, I.,
  {Staubert}, R., {Ferrigno}, C., \& {Klochkov}, D.,  2010a, \aap, 515, A10

\bibitem[\protect\astroncite{{Doroshenko} et~al.}{2010b}]{Doroshenko:2010}
{Doroshenko}, V., {Suchy}, S., {Santangelo}, A., {Staubert}, R., {Kreykenbohm},
  I., {Rothschild}, R., {Pottschmidt}, K., \& {Wilms}, J.,  2010b, \aap, 515,
  L1

\bibitem[\protect\astroncite{{Endo} et~al.}{2002}]{Endo:2002}
{Endo}, T., {Ishida}, M., {Masai}, K., {Kunieda}, H., {Inoue}, H., \& {Nagase},
  F.,  2002, \apj, 574, 879

\bibitem[\protect\astroncite{{Evangelista} et~al.}{2010}]{Evangelista:2010}
{Evangelista}, Y., et~al., 2010, \apj, 708, 1663

\bibitem[\protect\astroncite{{F{\"u}rst} et~al.}{2011a}]{Furst:2011}
{F{\"u}rst}, F., {Kreykenbohm}, I., {Suchy}, S., {Barrag{\'a}n}, L., {Wilms},
  J., {Rothschild}, R.~E., \& {Pottschmidt}, K.,  2011a, \aap, 525, A73

\bibitem[\protect\astroncite{{F{\"u}rst} et~al.}{2011b}]{Furst:2011b}
{F{\"u}rst}, F., {Suchy}, S., {Kreykenbohm}, I., {Barrag{\'a}n}, L., {Wilms},
  J., {Rothschild}, R.~E., \& {Pottschmidt}, K.,  2011b, \aap, in prep

\bibitem[\protect\astroncite{{G{\"o}{\u g}{\"u}{\c s}}, {Kreykenbohm} \&
  {Belloni}}{2011}]{Gogus:2011}
{G{\"o}{\u g}{\"u}{\c s}}, E., {Kreykenbohm}, I., \& {Belloni}, T.~M.,  2011,
  \aap, 525, L6+

\bibitem[\protect\astroncite{{Jones}, {Chetin} \& {Liller}}{1974}]{Jones:1974}
{Jones}, C.~A., {Chetin}, T., \& {Liller}, W.,  1974, \apjl, 190, L1

\bibitem[\protect\astroncite{{Kaastra} \& {Mewe}}{1993}]{Kaastra:1993}
{Kaastra}, J.~S., \& {Mewe}, R.,  1993, \aaps, 97, 443

\bibitem[\protect\astroncite{{Kaper}, {van der Meer} \&
  {Najarro}}{2006}]{Kaper:2006}
{Kaper}, L., {van der Meer}, A., \& {Najarro}, F.,  2006, \aap, 457, 595

\bibitem[\protect\astroncite{{Klochkov} et~al.}{2008}]{Klochkov:2008}
{Klochkov}, D., et~al., 2008, \aap, 482, 907

\bibitem[\protect\astroncite{{Klochkov} et~al.}{2011}]{Klochkov:2011}
{Klochkov}, D., {Staubert}, R., {Santangelo}, A., {Rothschild}, R.~E., \&
  {Ferrigno}, C.,  2011, \aap, submitted

\bibitem[\protect\astroncite{{Koh} et~al.}{1997}]{Koh:1997}
{Koh}, D.~T., et~al., 1997, \apj, 479, 933

\bibitem[\protect\astroncite{{Kreykenbohm} et~al.}{1999}]{Kreykenbohm:1999}
{Kreykenbohm}, I., {Kretschmar}, P., {Wilms}, J., {Staubert}, R., {Kendziorra},
  E., {Gruber}, D.~E., {Heindl}, W.~A., \& {Rothschild}, R.~E.,  1999, \aap,
  341, 141

\bibitem[\protect\astroncite{{Kreykenbohm} et~al.}{2004}]{Kreykenbohm:2004}
{Kreykenbohm}, I., {Wilms}, J., {Coburn}, W., {Kuster}, M., {Rothschild},
  R.~E., {Heindl}, W.~A., {Kretschmar}, P., \& {Staubert}, R.,  2004, \aap,
  427, 975

\bibitem[\protect\astroncite{{Kreykenbohm} et~al.}{2008}]{Kreykenbohm:2008}
{Kreykenbohm}, I., et~al., 2008, \aap, 492, 511

\bibitem[\protect\astroncite{{La Barbera} et~al.}{2003}]{La-barbera:2003}
{La Barbera}, A., {Santangelo}, A., {Orlandini}, M., \& {Segreto}, A.,  2003,
  \aap, 400, 993

\bibitem[\protect\astroncite{{La Barbera} et~al.}{2005}]{La-Barbera:2005}
{La Barbera}, A., {Segreto}, A., {Santangelo}, A., {Kreykenbohm}, I., \&
  {Orlandini}, M.,  2005, \aap, 438, 617

\bibitem[\protect\astroncite{{Leahy}}{2002}]{Leahy:2002}
{Leahy}, D.~A.,  2002, \aap, 391, 219

\bibitem[\protect\astroncite{{Leahy} \& {Kostka}}{2008}]{Leahy:2008}
{Leahy}, D.~A., \& {Kostka}, M.,  2008, \mnras, 384, 747

\bibitem[\protect\astroncite{{Lewin} et~al.}{1971}]{Lewin:1971}
{Lewin}, W.~H.~G., {McClintock}, J.~E., {Ryckman}, S.~G., \& {Smith}, W.~B.,
  1971, \apjl, 166, L69

\bibitem[\protect\astroncite{{Matsumoto} et~al.}{2006}]{Matsumoto:2006}
{Matsumoto}, H., et~al., 2006,
\newblock in Society of Photo-Optical Instrumentation Engineers, Proc. SPIE,
  Vol. 6266

\bibitem[\protect\astroncite{{McClintock}, {Ricker} \&
  {Lewin}}{1971}]{McClintock:1971}
{McClintock}, J.~E., {Ricker}, G.~R., \& {Lewin}, W.~H.~G.,  1971, \apjl, 166,
  L73

\bibitem[\protect\astroncite{{Meszaros} \& {Nagel}}{1985}]{Meszaros:1985}
{Meszaros}, P., \& {Nagel}, W.,  1985, \apj, 299, 138

\bibitem[\protect\astroncite{{Mihara}}{1995}]{Mihara:1995}
{Mihara}, T.,  1995,
\newblock Ph.D. thesis, Univ.~of Tokyo

\bibitem[\protect\astroncite{{Mihara} et~al.}{1990}]{Mihara:1990}
{Mihara}, T., {Makishima}, K., {Ohashi}, T., {Sakao}, T., \& {Tashiro}, M.,
  1990, \nat, 346, 250

\bibitem[\protect\astroncite{{Mihara} et~al.}{2007}]{Mihara:2007}
{Mihara}, T., et~al., 2007, Prog. Theor. Phys. Suppl., 169, 191

\bibitem[\protect\astroncite{{Mitsuda} et~al.}{2007}]{Mitsuda:2007}
{Mitsuda}, K., et~al., 2007, \pasj, 59, 1

\bibitem[\protect\astroncite{{Mowlavi} et~al.}{2006}]{Mowlavi:2006}
{Mowlavi}, N., et~al., 2006, \aap, 451, 187

\bibitem[\protect\astroncite{{Mukherjee} \& {Paul}}{2004}]{Mukherjee:2004}
{Mukherjee}, U., \& {Paul}, B.,  2004, \aap, 427, 567

\bibitem[\protect\astroncite{{Nakajima}, {Mihara} \&
  {Makishima}}{2010}]{Nakajima:2010}
{Nakajima}, M., {Mihara}, T., \& {Makishima}, K.,  2010, \apj, 710, 1755

\bibitem[\protect\astroncite{{Orlandini} et~al.}{2000}]{Orlandini:2000}
{Orlandini}, M., {dal Fiume}, D., {Frontera}, F., {Oosterbroek}, T., {Parmar},
  A.~N., {Santangelo}, A., \& {Segreto}, A.,  2000, Advances in Space Research,
  25, 417

\bibitem[\protect\astroncite{{Parkes} et~al.}{1980}]{Parkes:1980}
{Parkes}, G.~E. and {Culhane}, J.~L. and {Mason}, K.~O. and {Murdin}, P.~G.,
  1980, \mnras, 454, 872
  
\bibitem[\protect\astroncite{{Pravdo} et~al.}{1995}]{Pravdo:1995}
{Pravdo}, S.~H., {Day}, C.~S.~R., {Angelini}, L., {Harmon}, B.~A., {Yoshida},
  A., \& {Saraswat}, P.,  1995, \apj, 191, 547

\bibitem[\protect\astroncite{{Pravdo} \& {Ghosh}}{2001}]{Pravdo:2001}
{Pravdo}, S.~H., \& {Ghosh}, P.,  2001, \apj, 554, 383

\bibitem[\protect\astroncite{{Soong} et~al.}{1990}]{Soong:1990}
{Soong}, Y., {Gruber}, D.~E., {Peterson}, L.~E., \& {Rothschild}, R.~E.,  1990,
  \apj, 348, 641

\bibitem[\protect\astroncite{{Staubert} et~al.}{2007}]{Staubert:2007}
{Staubert}, R., {Shakura}, N.~I., {Postnov}, K., {Wilms}, J., {Rothschild},
  R.~E., {Coburn}, W., {Rodina}, L., \& {Klochkov}, D.,  2007, \aap, 465, L25

\bibitem[\protect\astroncite{{Suchy} et~al.}{2011}]{Suchy:2011}
{Suchy}, S., et~al., 2011, \apj, 733, 15

\bibitem[\protect\astroncite{{Takahashi} et~al.}{2007}]{Takahashi:2007}
{Takahashi}, T., et~al., 2007, \pasj, 59, 35

\bibitem[\protect\astroncite{{Tanaka}}{1986}]{Tanaka:1986}
{Tanaka}, Y.,  1986,
\newblock in Radiation Hydrodynamics in Stars and Compact Objects, ed.
  D.~Mihalas \& K.-H.~A.~Winkler, IAU Colloq. series 89, Vol. 255,  198

\bibitem[\protect\astroncite{{Tsygankov} et~al.}{2006}]{Tsygankov:2006}
{Tsygankov}, S.~S., {Lutovinov}, A.~A., {Churazov}, E.~M., \& {Sunyaev}, R.~A.,
   2006, \mnras, 371, 19

\bibitem[\protect\astroncite{{Tsygankov} et~al.}{2007}]{Tsygankov:2007}
{Tsygankov}, S.~S., {Lutovinov}, A.~A., {Churazov}, E.~M., \& {Sunyaev}, R.~A.,
   2007, Astron. Lett., 33, 368

\bibitem[\protect\astroncite{{Verner} et~al.}{1996}]{verner:1996}
{Verner}, D.~A., {Ferland}, G.~J., {Korista}, K.~T., \& {Yakovlev}, D.~G.,
  1996, \apj, 465, 487

\bibitem[\protect\astroncite{{Watanabe} et~al.}{2003}]{Watanabe:2003}
{Watanabe}, S., et~al., 2003, \apjl, 597, L37

\bibitem[\protect\astroncite{{White} et~al.}{1976}]{White:1976}
{White}, N.~E., {Mason}, K.~O., {Huckle}, H.~E., {Charles}, P.~A., \&
  {Sanford}, P.~W.,  1976, \apjl, 209, L119

\bibitem[\protect\astroncite{{White}, {Mason} \& {Sanford}}{1978}]{White:1978}
{White}, N.~E., {Mason}, K.~O., \& {Sanford}, P.~W.,  1978, \mnras, 184, 67P

\bibitem[\protect\astroncite{{Wilms}, {Allen} \& {McCray}}{2000}]{wilms:2000}
{Wilms}, J., {Allen}, A., \& {McCray}, R.,  2000, \apj, 542, 914

\end{thebibliography}
\end{document}